# REPORTING, REVIEWING, AND RESPONDING TO
# HARASSMENT ON TWITTER

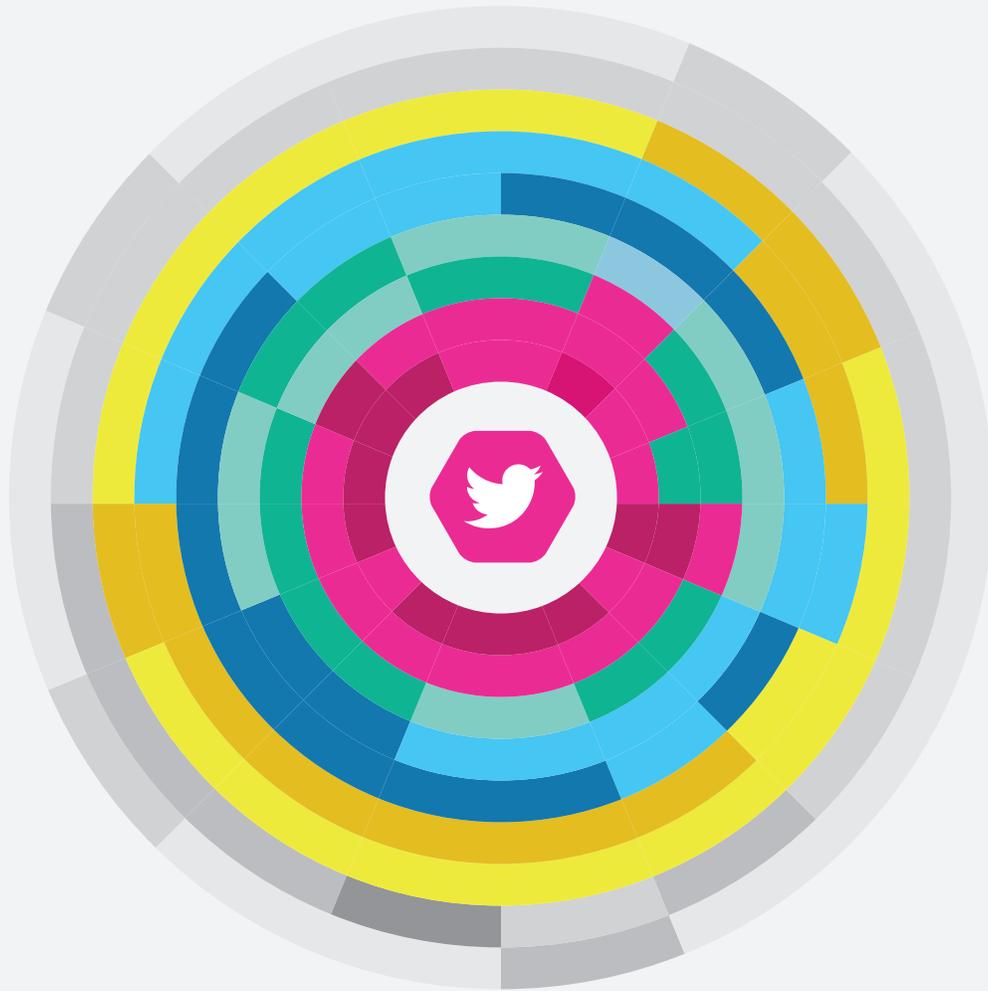

**FOR FURTHER INFORMATION
ON THIS REPORT, CONTACT:**
**Jamia Wilson,** Executive Director: Women, Action, and the Media
**wam@womenactionmedia.org**
**womenactionmedia.org**

# ABOUT THIS REPORT

This report was produced at the request of Women, Action, and the Media (WAM!). Aside from Jaclyn Friedman, founder and former executive director of WAM!, the authors of this document are academics from the fields of computational social science, anthropology, sociology, network science, and computer science. The academic authors conducted this analysis outside their academic institutions in an unpaid, volunteer capacity.

- **J. Nathan Matias,** M.A., M.S., Ph.D. Student in Computational Social Science

- **Amy Johnson,** M.A., Ph.D. Candidate in Anthropology

- **Whitney Erin Boesel,** M.A., Internet Researcher and Sociologist

- **Brian Keegan,** Ph.D., Computational Social Scientist

- **Jaclyn Friedman,** Founder and Strategic Advisor, Women, Action, & the Media

- **Charlie DeTar,** Ph.D., Software Engineer

This report was reviewed by five academic reviewers in a double-blind, revise-and-resubmit peer review process chaired by Zeynep Tufekci, Assistant Professor at the University of North Carolina, Chapel Hill. The authors are deeply grateful for the detailed feedback and high standards of these reviewers, whose efforts have led to a much clearer, stronger report.

## ABOUT WOMEN, ACTION & THE MEDIA

WAM! is an independent North American nonprofit dedicated to building a robust, effective, inclusive movement for gender justice in media, with chapters across the United States and Canada. For more information, visit www.womenactionmedia.org.

**What change is needed to make the Internet safer for free speech and equal participation?**
Building on its experience and the findings from this report, WAM! has prepared recommendations for Twitter and technology company decision makers, along with resources for people facing harassment and abuse. Learn more at **www.womenactionmedia.org/Twitter-Report**

# SUMMARY OF FINDINGS:

When people experience harassment online, from individual threats or invective to coordinated campaigns of harassment, they have the option to report the harassers and content to the platform where the harassment has occurred. Platforms then evaluate harassment reports against terms of use and other policies to decide whether to remove content or take action against the alleged harasser—or not. On Twitter, harassing accounts can be deleted entirely, suspended (with content made unavailable pending appeal or specific changes), or sent a warning. Some platforms, including Twitter and YouTube, grant "authorized reporters" or "trusted flaggers" special privileges to identify and report inappropriate content on behalf of others.

In November 2014, Twitter granted Women, Action, and the Media (WAM!) this authorized reporter status. From November 6–26 2014, WAM! took in reports of Twitter-based harassment, assessed them, and escalated reports as necessary to Twitter for special attention. WAM! used a special intake form to collect data and promised publicly to publish what it learned from the data it collected. In three weeks, WAM! reviewers assessed 811 incoming reports of harassment and escalated 161 reports to Twitter, ultimately seeing Twitter carry out 70 account suspensions, 18 warnings, and one deleted account. This document presents findings from this three-week project; it draws on both quantitative and qualitative methods.

Findings focus on the **people reporting and receiving harassment,** the **kinds of harassment** that were reported, **Twitter's response to harassment reports**, the process of **reviewing harassment reports**, and **challenges for harassment reporting processes**.

## WHO WAS REPORTING HARASSMENT?

People come to harassment reporting systems from a variety of situations and with a variety of needs and goals. From the project's qualitative data, ten profiles of people involved in harassment reporting emerge. These non-exclusive, composite profiles point to specific support needs during the reporting process.

Harassment reports are often submitted by *receivers of harassment*, people who are experiencing different stages or types of harassment:

**First-time receiver of harassment;** this person needs in-depth investigation and technical assistance, as well as listening and direction to external resources.

**People who are targets of chaining or suspension evasion,** forms of harassment involving a linked series of accounts. This person needs to be able to link harassment reports together and for Twitter to catch, prevent, and take action on harassers' chaining patterns.

**People experiencing dogpiling,** harassment from multiple accounts at the same time. This person needs to be able to take action on—and collect evidence on—multiple accounts at the same time.

**People experiencing harassment that crosses platforms;** this person needs to be directed to law enforcement and external support.

**People experiencing offensive but not harassing encounters online;** this person needs basic information about how harassment is determined, by Twitter and more broadly.

Not all reports are made by receivers of harassment—in fact, the majority of the harassment reports WAM! received came from other sources:

**Delegates,** authorized agents of a person receiving harassment**.** This person needs a simple but robust process for demonstrating authorization.

**Bystanders,** people who report harassment without the knowledge of the person being harassed**.** This person needs to know their report was taken seriously.

People sometimes use the harassment reporting process for functions other than reporting harassment:

**People who engage in false flagging** use the harassment reporting process to attempt to silence or intimidate an account.

**People who engage in report trolling** (pretending to having been harassed) intentionally expend valuable resources of reviewers.

**People who engage in spewing** use the reporting form to direct invective and threats at reviewers.

These composite profiles represent experiences detailed in this dataset; they may overlap at times and should not be taken as exclusive.

## Most Harassment Reports were On Behalf of Others

*% of WAM! reports that self-identified as the receiver of harassment, and the % of self-identified bystanders versus delegates.*

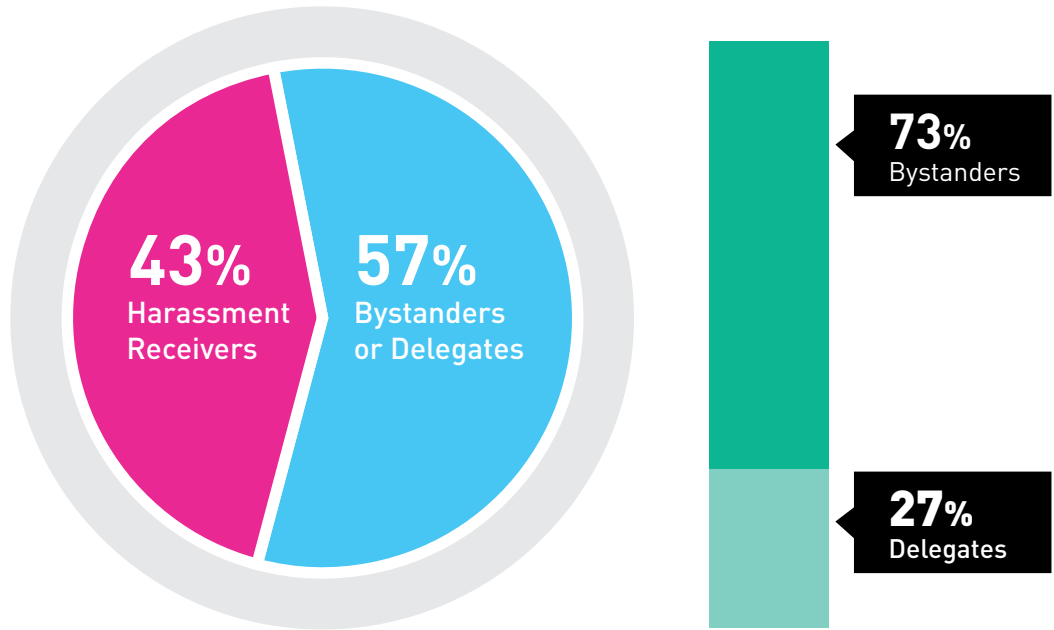

**43%** Harassment Receivers

**57%** Bystanders or Delegates

**73%** Bystanders

**27%** Delegates

SOURCE WAM! Harassment Reports, Nov. 6-26, 2014. n=317
**©Women, Action & the Media**

## WHAT KINDS OF HARASSMENT WERE REPORTED TO WAM!?

Among the 317 genuine harassments reports submitted to WAM!, hate speech and doxxing (releasing private information) were the most common, with 19% of cases representing reports that didn't fall neatly into any of the categories offered.

- Ongoing harassment was a concern in 29% of reports, where reporters mentioned that harassment started more than three weeks before the report.

- Most submitters claimed to have notified Twitter previously, with 67% claiming to have notified Twitter at least once about a case of harassment.

- Follow-up correspondence with receivers of harassment repeatedly mention that law enforcement claimed an inability to help or directed reporters back to corporations like Twitter.

- Harassment that was too complex to enter in a single radio button (19% of reports) comprised one of the greatest challenges for harassment receivers and WAM!, taking up large amounts of time and conversation to establish well.

- The GamerGate controversy, prompted by tensions over diversity in the videogame

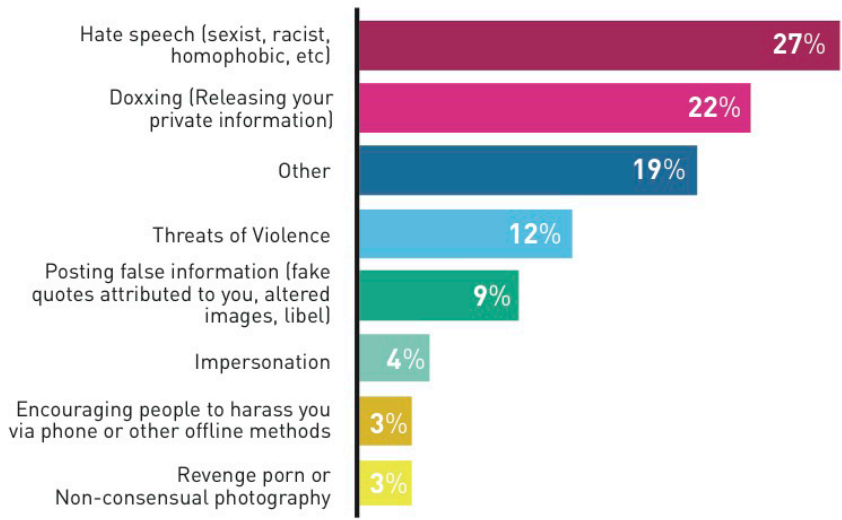

## Harassment Reports Focused on Hate Speech, Doxxing, Threats of Violence, and Other More Complex Cases

*Among reports submitted to WAM!, the % that self-identified with the following options of the WAM! reporting form.*

| | |
|---|---|
| Hate speech (sexist, racist, homophobic, etc) | **27**% |
| Doxxing (Releasing your private information) | **22**% |
| Other | **19**% |
| Threats of Violence | **12**% |
| Posting false information (fake quotes attributed to you, altered images, libel) | **9**% |
| Impersonation | **4**% |
| Encouraging people to harass you via phone or other offline methods | **3**% |
| Revenge porn or Non-consensual photography | **3**% |

SOURCE: WAM! Harassment Reports, Nov 6-26, 2014 n=317

**© Women, Action, and the Media**

industry and associated with some of the highest profile Twitter harassment in 2014, was a substantial but not primary source of harassment reports. Although WAM!'s reporting period occurred during the controversy, only 12% of the 512 alleged harassing accounts could be linked to GamerGate.

WAM! escalated 43% of genuine reports to Twitter, opening 161 tickets with Twitter's abuse reporting system on behalf of those who reported harassment to WAM, escalating over half of incoming reports of revenge porn, impersonation, false information, and doxxing.

### HOW DID TWITTER RESPOND TO ESCALATED REPORTS?
WAM! collected data on the process and outcome of all 161 tickets opened with Twitter in the three week monitoring period.

**In 55% of cases, Twitter took action to delete, suspend, or warn the alleged harassing account.** Most of Twitter's actions against alleged harassers were associated with reports of hate speech, threats of violence, and nonconsensual photography.

Was Twitter more likely to take action on some kinds of harassment and not others? In a logistic regression model, **the probability of Twitter taking action on reports of doxxing was 20 percentage points lower than tickets involving threats of violence**, in cases where WAM! recorded an assessment risk, an odds ratio of 0.32. This is likely due to the common practice of 'tweet and delete,' in which harassers temporarily post private, personal information and remove the content before it can be reported and investigated by Twitter.

Does Twitter favor longstanding or popular accounts in its decisions to take action, with

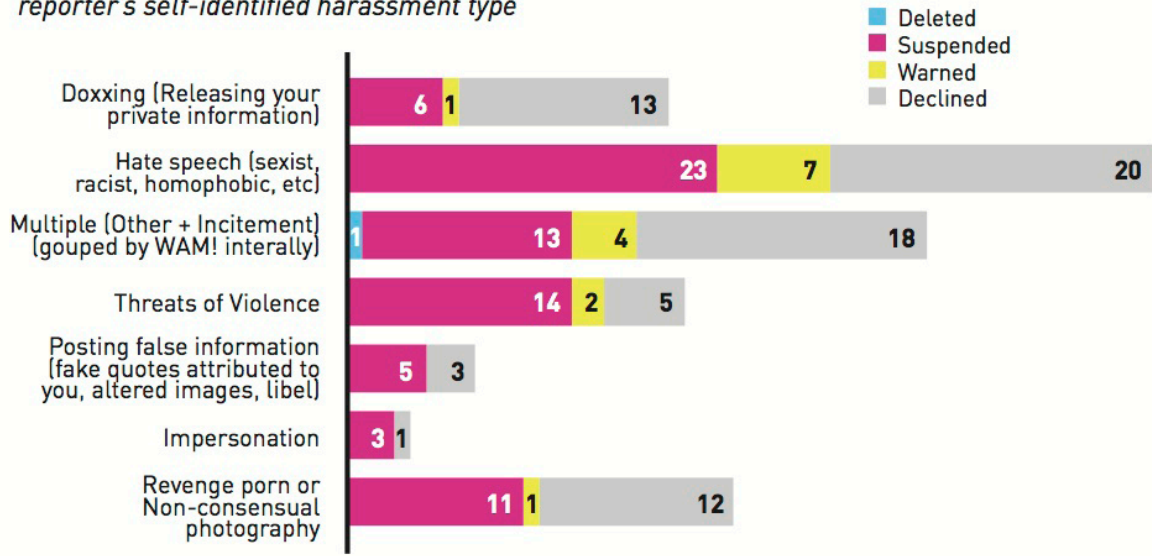

## Twitter's Actions Toward Allegedly Harassing Accounts In Response to Reports Escalated by WAM!

*Number of actions by Twitter to delete, suspend, warn, or decline action, in response to reports escalated by WAM!, grouped by the reporter's self-identified harassment type*

Legend:
- Deleted
- Suspended
- Warned
- Declined

| Harassment type | Suspended | Warned | Declined |
|---|---|---|---|
| Doxxing (Releasing your private information) | 6 | 1 | 13 |
| Hate speech (sexist, racist, homophobic, etc) | 23 | 7 | 20 |
| Multiple (Other + Incitement) (gouped by WAM! interally) | 13 | 4 | 18 |
| Threats of Violence | 14 | 2 | 5 |
| Posting false information (fake quotes attributed to you, altered images, libel) | 5 | | 3 |
| Impersonation | 3 | | 1 |
| Revenge porn or Non-consensual photography | 11 | 1 | 12 |

SOURCE: Composite data on WAM! Harassment Reports, Nov 6-26, 2014, linked with WAM! Ticketing system and coded emails of Twitter's responses to WAM!'s escalated tickets. In some cases, multiple harassment reports (and therefore multiple types of harassment) are sometimes associated with a single Twitter action.  n=161

**© Women, Action, and the Media**

regard to either reporters of harassment or alleged harassers? Within the limitations of the data available, **an analysis of Twitter's actions did not find evidence of favoritism by Twitter** based on the age of an account or number of followers.

### THE AUTHORIZED REPORTER RELATIONSHIP

The authorized reporter relationship can help both individual reporters and platforms navigate the challenges of reporting, assessing, and responding to harassment. Authorized reporters can serve as allies—advocates who not only escalate reports of harassment but offer reporters and receivers of harassment valuable understanding, emotional support, and external resources. However, the authorized reporter relationship also carries complications:

- Notably, **WAM!'s status as an authorized reporter was widely misunderstood.** People reporting harassment to WAM!, the news media, and occasionally even Twitter staff demonstrated considerable confusion about the nature of the relationship.

- The authorized reporter relationship is unstable: **The platform determines who is granted authorized reporter status, how the relationship works in practice, and whether or not it continues.**

### THE WORK OF REVIEWING HARASSMENT

During the three-week project, WAM! reviewers assessed an average of 30 incoming reports per day, responded personally to 17 reports per day, wrote an average of 58 messages of discussion per day, and exchanged messages with Twitter at least 9 times a day. During this time period, WAM! reviewers wrote nearly 60,000 words in exchanges with receivers of harassment and each other.

🔍 Reviewing reports with **attention to the experience of harassment requires in-depth communication**. People experiencing harassment often have complex situations with needs that only become clear through multiple exchanges.

🔍 The reporting process is an opportunity to establish trust and listen. **Processes optimized solely for stopping harassment are unlikely to address the larger impact of the harassment on the targeted user**.

🔍 **Reviewing reports can have serious mental health consequences for reviewers**; genuine evidence may be triggering and false-flag reporters may submit harmful material in malicious reports.

## THE PROBLEM OF EVIDENCE

The challenges of proving harassment are often framed as challenges of context and interpretation. Experiences from the WAM! project emphasize that the mode or format of evidence also has serious consequences.

■ Twitter currently requires URLs and rejects screenshots as evidence; consequently, Twitter's review process doesn't address **'tweet and delete' harassment, which often involves doxxing**. While Twitter updated its reporting system in February 2015 to accept reports of doxxing,[1] there have been no public changes with regard to the evidence it accepts for harassment reports.

■ Twitter's default URL requirement makes it **complicated to report harassment that is not associated with a URL**, such as exposure to violent or pornographic profile images or usernames via follower/favorite notifications.

When platforms decide to act on harassment reports by removing content or suspending accounts, their actions can also affect the evidence available for other channels of reporting.

■ **Account suspension can impede people seeking to report harassment to law enforcement,** particularly when done without public explanation. To help address this, in March 2015 Twitter began offering an option for reporters to receive an emailed record of the report they submitted.[2]

# INTRODUCTION

Online harassment has been acknowledged as a problem from the early days of the internet.[3,4,5] In 2014, a series of high profile cases pushed the issue to media prominence. Zelda Williams, daughter of actor Robin Williams, shut down her Twitter account for several weeks after large scale harassment followed her father's death, prompting the company to revisit its policies.[6,7] The GamerGate controversy—over tensions linked to growing gender and cultural diversity in videogames—saw numerous cases of online harassment. Harassment on Twitter associated with the controversy led game critic Anita Sarkeesian,[8] game developer Zoe Quinn,[9] and game developer Brianna Wu[10] to flee their homes and cancel appearances after harassers published their physical addresses (a practice called doxxing), threatened them, and incited others to violence against them. In October 2014, the Pew Research Center published a study of the experience of online harassment in the United States: 40% of Americans have experienced harassment in some form.[11]

Debates about online harassment overlap with issues of free speech. In 2014, the U.S. Supreme Court heard arguments in the case of Elonis vs the United States, debating the limits of speech rights online in cases of threats of violence.[12] Harassment can also be a means to silence the speech of others, especially women, an argument raised in a January 2014 *Pacific Standard* article by Amanda Hess[13] and in an April 2015 *Washington Post* article by Twitter general counsel Vijayada Gadde.[14]

Many platforms have systems for reporting, reviewing, and responding to harassment. This can involve outsourcing the work to third party companies; in one company, over 100,000 people reportedly review and respond to flagged activity.[15,16] Responses can vary from removal of content, to action against a user, to a decision not to intervene. The policies, systems, and staff involved in handling these reports received greater public scrutiny in 2014.[17,18] Twitter has reportedly tripled the support team handling abuse reports.[19]

Some platforms, including Twitter and YouTube, grant "authorized reporters" or "trusted flaggers"[20] special privileges to identify and report inappropriate content on behalf of others. In November 2014, the advocacy organization

## How WAM! Handled Reports of Harassment

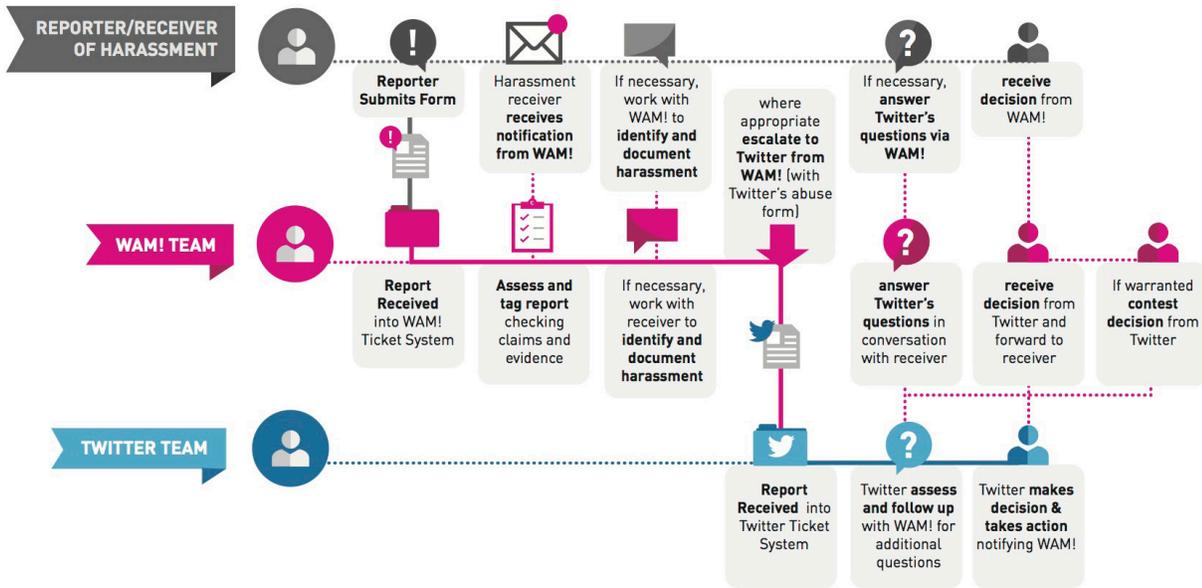

*Figure 1: How WAM! handled reports of harassment*

Women, Action, and the Media (WAM!) was granted authorized reporter status by Twitter.[21] This status enabled WAM! to take reports of harassment on Twitter, assess them, and escalate reports as necessary to Twitter, where the company would give the reports special attention. WAM! created a special intake form to receive harassment reports and collect data for the first three weeks, promising publicly to publish what it learned from the data it collected.

Although the authorized reporter role was new for WAM!, the advocacy organization has a history of engaging with social media platforms. As a North American nonprofit focused on gender justice in the media, WAM! is a network of grassroots community groups that have historically focused on broadening inclusion and speech for marginalized groups in the media. In 2013, WAM! participated in a campaign with the Everyday Sexism Project and author/activist Soraya Chemaly that successfully prompted Facebook to improve its content moderation policies towards content targeting women.[22,23] WAM!'s engagement with Twitter through its authorized reporter status was an extension of that earlier work.

The project and WAM!'s new status with Twitter were publicized widely in the media. Indeed, WAM!'s Twitter reporting project was the most widely covered story about harassment on Twitter in 2014, generating more than 204 stories in 21 countries. Though coverage of the project was largely positive, its structure wasn't well understood: despite the efforts of WAM! and Twitter, media repeatedly mischaracterized the authorized reporting status as a "partnership" or "collaboration." A more extended analysis of media coverage of the project can be found in Appendix 2: Media Coverage.

In its simplest form, Twitter's authorized reporter status involves two key aspects: The ability of a third-party organization, vetted by Twitter, to report on behalf of an individual, and a different prioritizing of those reports by Twitter.[24] WAM! is not the only organization that holds authorized reporter status with Twitter, and though not widely publicized, other platforms utilize similar relationships. In most ways, the reporting process for an authorized reporter is similar to that of an individual reporting harassment directly to Twitter. Thus, WAM! staff filled out the same reporting forms that an individual would and engaged in what seems to be similar follow-up correspondence with Twitter staff.

Beginning on 6 November 2014, WAM! invited members of the public to submit reports of harassment on Twitter through their form located on the WAM! website. Over the course of the subsequent three-week project, WAM! reviewers assessed 811 incoming reports of harassment and escalated 161 reports to Twitter, ultimately seeing Twitter carry out 70 account suspensions, 18 warnings, and one account deletion.

After the project concluded, WAM! invited a group of researchers, to analyze the data they had collected on harassment and review their larger authorized reporting process. This included a variety of types of evidence:

- 811 initial reports of harassment received via the WAM! intake form

- 640 tickets within WAM!s internal system for responding to reports, including correspondence between WAM! and reporters of harassment and internal WAM! discussion of those reports

- 185 messages associated with reports escalated to Twitter, representing 154 tickets opened up by WAM! with Twitter.

- internal WAM! documentation for the project

- personal accounts of the project and its design from participants

In the course of analysis, additional evidence was also consulted, including Twitter's reporting tools and policies, public Twitter accounts, and media articles about the WAM! project. This document presents the findings of that analysis, drawing on a variety of qualitative and quantitative methods, including discourse analysis and statistical analysis.

At the heart of all of this analysis resides a focus on the human experience of reporting harassment and reviewing harassment. This document is intended as a transparent, comprehensive resource that companies, law enforcement, advocates, and scholars can use to improve safety and support for people facing harassment online.

There are important limitations to the findings reported here. Notable among these are limitations in representativeness of the data in terms of volume and language, perceptions of WAM! as an entity with a specific identity, and WAM!'s decision to decline to pursue a number of reports that didn't show evidence of "gendered harassment." Detailed descriptions of data, methods, and limitations can be found in Appendix 1: Data, Methods & Limitations.

Meanwhile, Twitter has begun to change the way it handles harassment. In December 2014[25] Twitter launched a major update to its reporting tools, streamlining the process and allowing bystander reporting. In February[26] and March[27] 2015 Twitter introduced additional changes, with implications for doxxing and reporting to law enforcement, among others. While positive steps forward, for the most part these updates have little bearing on the data and analysis included in this report. In the few instances where the report

---

details problems that Twitter updates have addressed, relevant Twitter updates are noted.

## ORGANIZATION & TERMINOLOGY

This document is divided into two main parts: Part 1 examines who reported harassment to WAM!, what was reported, and how Twitter responded to WAM!'s escalated reports. Part 2 explores the work of reviewing harassment through the lens of the WAM! project: the authorized reporter relationship, the design of the WAM! reporting tool, the labor involved in reviewing and responding; it closes with a discussion of the problems of evidence that the project made visible. Appendixes include information on methods, media coverage, and the WAM! Twitter reporting tool used to collect initial reports.

At times the terms discussing the various people and actions involved in the project can be confusing. Often this is because *reporting*, *reviewing*, and *responding*—the three actions that ground this project—are performed by different actors: individuals report to WAM! (and sometimes Twitter as well) while WAM!, as an authorized reporter, reports to Twitter; both WAM! and Twitter engage in review processes; and both WAM! and Twitter respond to the reports they receive.

A second area of complexity revolves around differences in the actors who submit reports. The category of individual reporters—often just 'reporters'—includes anyone who submitted a report through the WAM! intake form. Sometimes these reporters were also the receivers of harassment. In the majority of cases, however, reports were submitted on the behalf of others, either by their delegates (authorized agents) or by bystanders, who did so without the awareness of the receiver of harassment.

Finally, it's difficult to discuss harassment without tangling with assumptions about power, legality, and context. The authors strive to treat all individuals with respect and humanity. Thus, where possible, in this document the main actors in harassment interactions are primarily referred to as 'receiver of harassment' and

'alleged harasser.' These terms are ungainly and imperfect, but preferable to more loaded options like 'victim' and 'attacker.' Judging communications as harassment or not falls outside the scope of this analysis. Rather, the analysis presented here focuses on acts of reporting, reviewing, and responding.

## ETHICS

WAM! publicly promised to use the project data for research when the project was announced. On the project's intake form, rollover explanations indicated data was being collected for research. And in cases where Twitter declined to take action, the email template WAM! sent to reporters referenced the collection of data for research purposes. WAM! did not, however, explicitly request consent to use this data for research.

WAM! invited the authors of this report to analyze data from the WAM! project only after that data had already been collected. Authors were asked to sign nondisclosure agreements to protect the privacy of the data. For the academics among the authors, this analysis was performed outside formal affiliations with home institutions in an unpaid, volunteer capacity. All academics among the authors have taken institutional ethics training for research with human subjects.

Harassment and reporting harassment are sensitive, controversial topics. The authors have undertaken data analysis with the highest possible regard for safety and privacy. With WAM!'s full cooperation, the following measures have been taken to ensure the privacy of people who trusted WAM! with their reports:

all data has been securely stored, accessed only via encrypted connection all authors have signed legal agreements binding authors to respect the privacy of this information no personally identifying information, including names, email addresses, phone numbers, Twitter account names, quotations from tweets, or messages exchanged with WAM!, are revealed in this report no personal information has or will be shared outside of WAM!, except for the purposes of escalating reports for action by Twitter no other personal information has or will be shared between WAM! and Twitter beyond reports escalated by WAM! to Twitter as part of its authorized reporter relationship

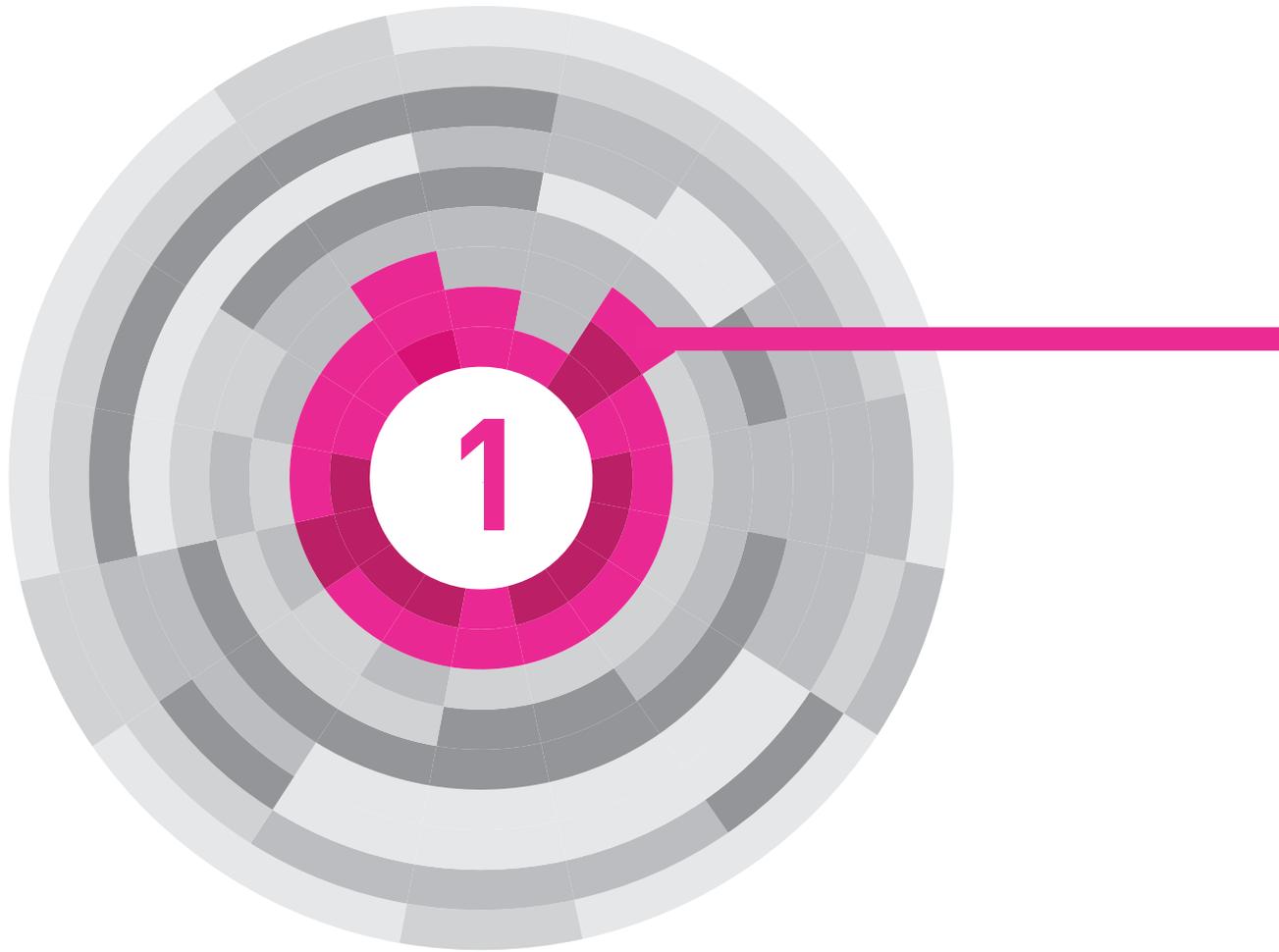

# PART ONE:
# HARASSMENT REPORTERS, REPORTS, AND OUTCOME

# 1.1

## WHO REPORTED HARASSMENT?

Who came to WAM! and what did they need?[28] People came to the reporting process with different experiences of harassment in terms of stage and type, some appalled and in shock from their first encounter with harassment, some exhausted and frustrated from dealing with extended campaigns of harassment. People came in different roles, some reporting on behalf of themselves, some for others with or without their knowledge. People came to report harassment—but people also came to intimidate, troll, and spew.

The following ten profiles illustrate a variety of the situations, needs, and goals people articulated in the context of reporting harassment to WAM!. The profiles draw on the rich details of problems, contexts, and histories shared in reports to WAM! and subsequent correspondence. They are constructed as composites to highlight similarities visible in the data through qualitative analysis, as well as to protect the anonymity of the individuals involved. The profiles emphasize the reporting process and support needs because the information available through the WAM! project is inextricably intertwined with the reporting process.

These ten profiles offer broader insights into reporting experiences and identify avenues for improvement. However, they are composites drawn from the WAM! project. As such, they do not describe every experience of harassment or reporting harassment; details may vary. These profiles also are not exclusive: people may face

28. The WAM! Twitter Reporting Tool did not query reporters about traditional demographic variables like sex, race, age, etc.

more than one challenge at a time, and may face others not detailed here.

It is important to note that these profiles present the support challenges of people who are already reporting harassment. These are people who have invested in the reporting process, indicating a desire to continue to engage with Twitter. The following profiles do not address the challenges of people who are *not* reporting harassment, whether because they have chosen to leave the Twitter platform or for some other reason.

## CHALLENGES FACED BY PEOPLE REPORTING HARASSMENT

Harassment reports are often submitted by *receivers of harassment*, people who are experiencing different stages or types of harassment:

**The first-time receiver:** This is the first time this person has had to deal with harassment on social media and the person is in shock. They may mistake WAM! for Twitter, Twitter for Facebook, and fill out the reporting form inaccurately. They're used to copying and pasting, and know how to screenshot, but need to be taught how to find a tweet's URL. They aren't familiar with harassment tactics or reporting harassment, and don't know how to report harassment directly from a tweet or from an account profile. By the time they recover from their experience enough to report harassment, tweets and images may have been deleted or protected. They're likely to assume that Twitter is aware of the specifics of all content on its platform and that its review team has investigatory capabilities that extend beyond the platform. **This person needs in-depth investigation and technical assistance, as well as listening and direction to external resources.**

**The target of chaining:** This person is being harassed by a single individual or few individuals, where the harassing accounts are regularly suspended only to be replaced by new ones run by the same harassers (suspension evasion). The person being harassed is receiving direct threats, insults, or graphic materials. These are typically easily recognized as violations of Twitter's terms of service and rules, and reporting them leads to account suspension. Harassers then open new accounts and begin their practices again. Key to this process is signaling to the target of harassment that a new account is related to one that was suspended. This is often achieved through use of a similar account name, recognizable profile details, or explicit boasting of continued presence despite suspensions. This signaling is a critical point of intimidation—but also opens a potential point of intervention. **A person being harassed through chaining needs to be able to link reports together, and they need Twitter to catch, prevent, and take action on that signaling.**

**The dogpiled:** This person is being harassed by many accounts concurrently, often as part of a campaign or coordinated attack. They're receiving—and receiving and receiving—threats and insults. This is high volume harassment. Consequently, filing a new report for each harassing account becomes an extension of the harassment, something harassers know. **Their unit of harassment isn't an individual account or an individual tweet, it's a wave of @mentions occurring during a particular time period. This person needs to be able to take action on—and collect evidence on—multiple accounts at the same time.**

**The target of multiplatform harassment:** This person is experiencing harassment on Twitter as part of a larger trajectory of harassment that moves and builds across multiple platforms. This involves threats, insults, and other unwanted contact directed toward them via channels such as social media accounts, email, blogs, professional presence (online and otherwise), and telephone. Harassment may come from a single individual or a group. Harassers follow, friend, and contact people this person knows, and use the person's online presence to stalk and/or publicly humiliate them. Harassers may also be impersonating the individual on other platforms and inciting others to attack them. This is a multi-headed hydra; individual platforms can only address the heads that they see, and chopping off those heads may obscure the growth of new ones elsewhere. **This person needs to be directed to law enforcement and external support.**

**The offended (but not harassed):** This person's sensibilities are offended. They have a public persona, but don't see themselves as a public figure. They may be receiving belittling reviews or insults with regard to professional choices or a business they run. If parody accounts are involved, they may see such accounts as impersonation or identity theft. They define harassment in conjunction with their personal irritation and give only minimal consideration to freedom of expression. They are unlikely to be familiar with legal or institutional processes for determining harassment. **This person needs basic information about how harassment is determined—by Twitter but also more broadly—and direction to external resources that provide more nuanced discussion and options for response that aren't limited to removing content or accounts.**

Not all reports are made by receivers of harassment—in fact, the majority of the harassment reports WAM! received came from other sources. Two profiles highlight the characteristics and support needs of people reporting harassment on behalf of others:

**The delegate:** This person is an authorized agent for the person experiencing the harassment. They may be an attorney hired to handle the case or a loved one insulating the person from further stressful or triggering interaction. The reporting they do is done with the awareness of the person experiencing the harassment. However, they face a challenge right from the start: they have to demonstrate that they have the authority to act on behalf of the person experiencing harassment. This is to be expected—even demanded. **This person needs a simple but robust process for demonstrating authorization. Further, they need information about the reporting process that they can communicate to the person experiencing the harassment that won't expose that person to additional distress.** This could be information about timeframe, number of review stages, staff involved in the review process, etc**.**

**The bystander:** This person is concerned for someone else's well-being and willing to take action behind the scenes. They're seeing someone getting harassed—probably doxxed, encouraged to kill themselves, or dogpiled—and are alarmed. The harassment is ongoing and potential consequences seem dire. They want someone with authority and/or administrative powers to step in. They're reporting the harassment without the knowledge of the person being harassed, and aren't authorized to act as that person's agent. **This person needs an assessment and potential intervention, and they need to know that there was assessment/intervention.**

People also sometimes use the harassment reporting process for functions other than reporting harassment. During the WAM! project, a number of people interacted with the WAM reporting tool maliciously. Such bad actors may be individuals acting on their own or members of a group; they may report manually or by script. These individuals are aware of and engaging with the reporting process. While their actions highlight the need to distinguish genuine from false reports, they also offer opportunities for education and connection:

**The false flagger:** This person falsely reports an account for harassment. This person intentionally tries to use Twitter's policies and the complexity of determining harassment to silence an account. This person may also report accounts falsely to draw reviewers' attention to themselves and their stance on issues under contention, often as an act of intimidation or warning. They may provide inaccurate contact details.

**The report troll:** This person performs a character, pretending to have been harassed. Their reports are marked by reductive narratives and stereotypical expressions, and often contain internal indicators such as word play, name choices, etc., that point to the performance aspect. They may provide functioning contact details under their character's persona in order to lengthen the performance.

**The spewer:** This person uses the reporting form to direct invective and threats at reviewers. This may be intended as anything from criticism to harm. They are unlikely to provide accurate contact information.

# 1.2

## HARASSMENT REPORTED TO WAM!

Between November 6 and November 26, WAM! received 317 reports that reviewers judged genuine, ranging from a minimum of 6 reports received on the 22nd to a maximum of 45 reports on the 20th.

### WHO SUBMITTED REPORTS?

**Bystanders and delegates submitted the majority of** harassment reports. 57% of all reports were on behalf of someone else, while 43% were from people claiming to be the person harassed. Among the 180 reports on behalf of others, 73% of self-identified bystanders claimed that the targeted person was aware of the report, while 27% of reports were made by self-identified delegates, claiming that the harassment receiver was aware of their report.

### PREVIOUS REPORTING TO TWITTER

**Most submitters claimed to have notified Twitter previously**. Out of 317 reports, 67% mentioned notifying Twitter previously at least once about the case of harassment; 18% of

## Most Harassment Reports were On Behalf of Others

*% of WAM! reports that self-identified as the receiver of harassment and the % of self-identified bystanders versus delegates.*

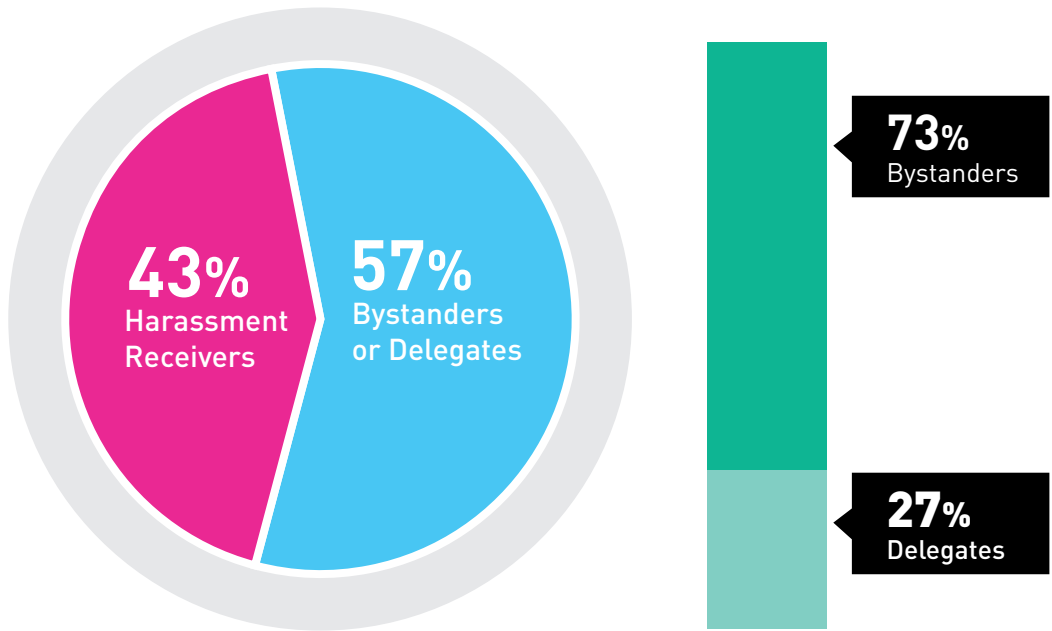

**43%** Harassment Receivers

**57%** Bystanders or Delegates

**73%** Bystanders

**27%** Delegates

SOURCE WAM! Harassment Reports, Nov. 6-26, 2014. n=317
**©Women, Action & the Media**

## Most Alleged Harassers Were Mentioned in a Single Report

*Number of reports containing alleged harassers*

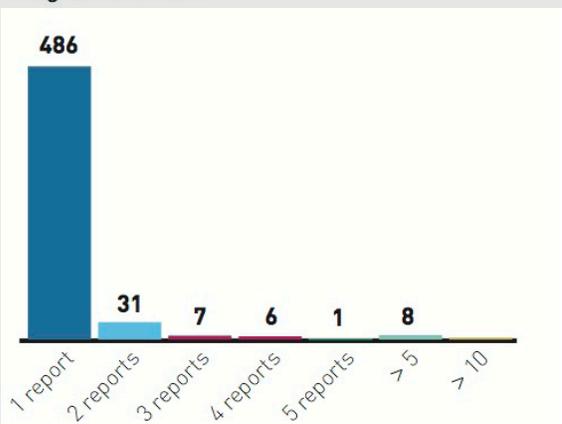

SOURCE: Unique Twitter accounts mentioned in WAM! Harassment Reports Nov 6–26,2014. n=538

© Women, Action, and the Media

## Most Reports Mentioned a Single Alleged Harasser

*Number of accounts mentioned per report*

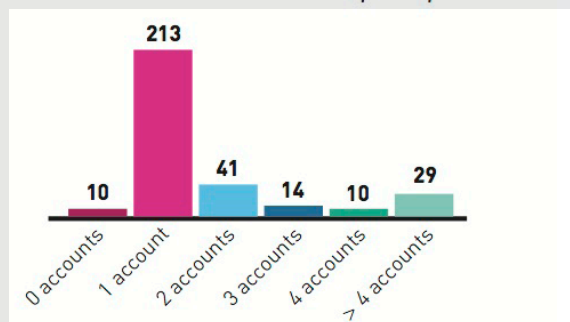

SOURCE: WAM! incoming Harassment Reports Nov 6–26, 2014. n=317

© Women, Action, and the Media

submitters claimed they had notified Twitter at least five times previously.

Some submitters used this field to express their frustration with Twitter, adding very high numbers, such as 9001[29] in this field. For this reason, no average is offered.[30]

### NUMBER OF ALLEGEDLY HARASSING ACCOUNTS PER REPORT

In an analysis of Twitter accounts mentioned in initial reports to WAM!, **most reports (67%) mentioned one allegedly harassing Twitter account**; 13% of reports mentioned two accounts and 9% mentioned more than four accounts.[31]

**Most allegedly harassing accounts were reported only once.** Submitters mentioned 538 unique accounts, out of which 90% appear only in one report. 10% of reported accounts (52) appeared in more than one report, with 8 accounts reported five times or more and 3 accounts reported ten times or more.

## Most Alleged Harassers Were Unconnected With GamerGate

*% of alleged harassers in the ggautoblocker list*

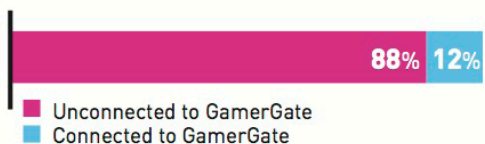

■ Unconnected to GamerGate
■ Connected to GamerGate

SOURCE: WAM! incoming Harassment Reports Nov 6–26, 2014. n=538 and Randi Harper's "Good Game Auto Blocker," Dec 29 2014.

© Women, Action, and the Media

.................................
29. A possible reference to the popular "Over 9000" Internet meme
30. Our counts of previous reporting to Twitter filter out all values above 200, to account for entries that most obviously take this approach.
31. In some cases, further accounts were discussed in subsequent communications with WAM!. In some analyses, we include those accounts, but we exclude them here since we have not coded these accounts as harassment receivers or allegedly harassing accounts.

The GamerGate controversy, notable for its connection with harassment on Twitter, was ongoing at the time of data collection. To check the influence of GamerGate on these findings, the authors investigated the proportion of WAM! reports that could be linked to GamerGate.[32] Reports to WAM! constitute a much wider range of harassment than the GamerGate controversy alone: 88% of allegedly harassing accounts (n=538) were not linked with GamerGate.

### NUMBER OF HARASSMENT RECEIVERS PER REPORT

**Most reports mentioned only one account receiving harassment**. Out of 317 reports, 85% reported a single harassment receiver; 13% mentioned more than one harassment receiver.

**Most harassment receivers were reported only once**. Among the 179 unique accounts that were reported as receiving harassment, most (73%) were reported only once, 29% were reported more than once, and 6% were reported five or more times.

Consistent with the high proportion of bystander reports, among the 179 unique harassment receivers associated with reports to WAM!, **more than a quarter of harassment receivers (28%) were mentioned in multiple reports** (51). Four accounts were reported ten times or more.

### WERE ALLEGED HARASSING ACCOUNTS ALSO REPORTED RECEIVERS OF HARASSMENT?

Among the entire set of Twitter accounts appearing in the WAM! reports, 27 accounts were mentioned in some reports as receivers of harassment and in others as allegedly harassing accounts. This constitutes 5% of unique allegedly harassing accounts and 15% of unique reported receivers of harassment. Why might this be? In some cases, receivers of harassment may also be engaging in activity that might constitute harassment. In other cases, receivers of harassment may be subject

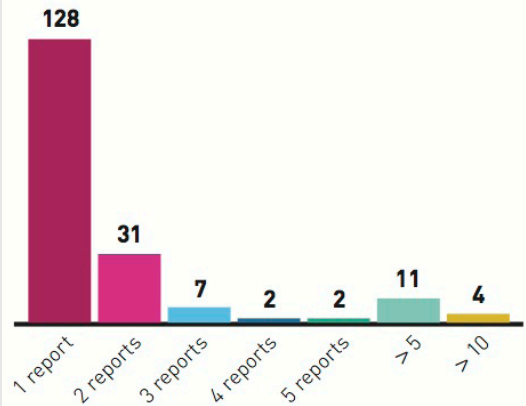

## Most Harassment Receivers Were Reported Only Once

*Number of reports containing harassment receivers*

SOURCE: Unique Twitter accounts mentioned in WAM! Harassment Reports Nov 6-26, 2014. n=179

**© Women, Action, and the Media**

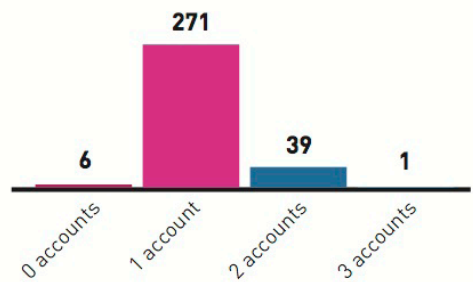

## Most Reports Mentioned a Single Receiver of Harassment

*Number of accounts mentioned as harassment receivers per report*

SOURCE: WAM! incoming Harassment Reports Nov 6-26, 2014. n=317

**© Women, Action, and the Media**

---

32. For more details, See Appendix 1.5: Methods: GamerGate Analysis

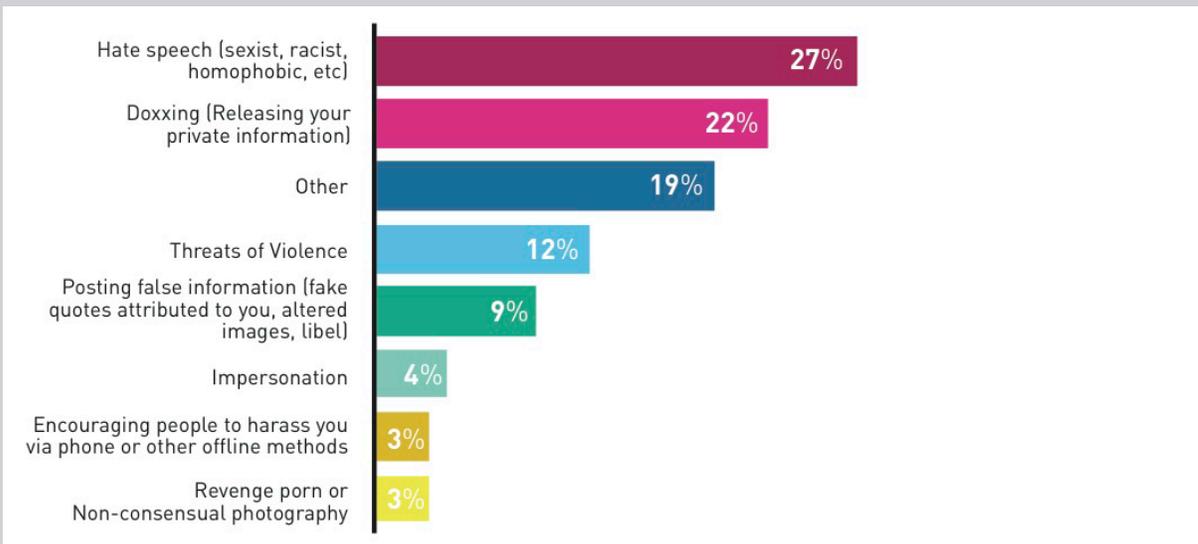

**Harassment Reports Focused on Hate Speech, Doxxing, Threats of Violence, and Other More Complex Cases**

*Among reports submitted to WAM!, the % that self-identified with the following options of the WAM! reporting form.*

SOURCE: WAM! Harassment Reports, Nov 6-26, 2014 n=317
**© Women, Action, and the Media**

to 'false flagging' campaigns that attempt to silence the harassment receiver through bad faith reports.

## KINDS OF HARASSMENT REPORTED

The WAM! reporting form asked reporters to categorize the type of harassment they were reporting, drawing on eight preassigned categories. The associated radio buttons of the form allowed reporters to select only one of these eight. This aspect of form design significantly affected the categorization of harassment, as revealed by responses in the free text of the 'Other' category, answers to the subsequent question *Please describe in detail the harassment you are receiving*, and later correspondence.

## HOW LONG HAD THE HARASSMENT BEEN OCCURRING?

Reporters were asked what date harassment had begun. 81 reports (26%) noted that harassment began the same day as the report, while 57%

of reports noted harassment had begun within the prior seven days (182). **Ongoing harassment was a concern for 29% of reporters (92), who mentioned that harassment had begun more than three weeks before submitting a report**.[33] In 14% of cases (45), harassment had reportedly begun more than six months earlier[34].

## IS HARASSMENT LIMITED TO TWITTER?

Reporters were asked if the harassment was occurring on multiple platforms. 54 reports (17%) mention harassment taking place on multiple platforms. Reporters were then asked to list the platforms, with 52 responses. Of these, 24 (46%)

...............................

33. In this analysis, we filter out four reports suggesting a harassment start date higher than 18.5 years ago. The largest duration included in this analysis is 5 years 10 months, with most falling under 5 years 6 months.
34. Other reporters included this information in further conversations with WAM!. That information is not included in this analysis, which is limited to answers on the form.

referenced more than one additional platform or channel and 27 (52%) mentioned only one additional platform or channel.[35]

Among those who mentioned harassment outside of Twitter, this took two forms. In many cases, reporters mentioned harassment that threatened individuals directly through email, personal blogs, telephone calls, and physical channels. Reported harassment also took advantage of the publicness of platforms like Facebook, Instagram, and YouTube to humiliate or embarrass targets in front of others.

Across all 54 reports of harassment outside Twitter, Facebook was often indicated, mentioned specifically in 17 responses (31%).[36] Email was mentioned in 9 reports (17%). Instagram, Tumblr, YouTube, Reddit, and blog (as a general category) all received several mentions. There were also 3 mentions of telephone calls and 2 of physical harassment.

**This data points to the limitations of addressing harassment through a single platform like Twitter**. Many of the people who were harassed on multiple platforms had already contacted law enforcement. While some reporters describe law enforcement taking responsive action, **correspondence repeatedly mentions law enforcement claiming an inability to help or directing reporters back to organizations like Twitter.**

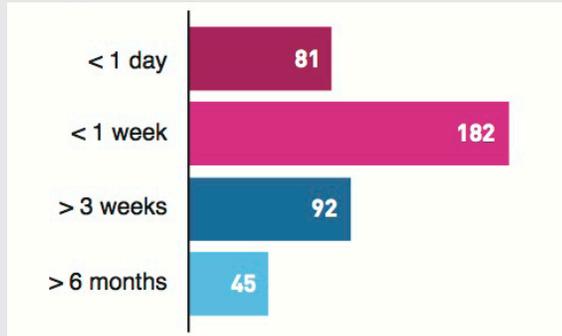

### Ongoing Harassment Was a Concern For Many Reporters

*Grouped answers to the form question "How many days since harassment started?"*

| | |
|---|---|
| < 1 day | 81 |
| < 1 week | 182 |
| > 3 weeks | 92 |
| > 6 months | 45 |

SOURCE: WAM! Harassment Reports. Nov 6-26, 2014. Cases greater than 65 years are filter out n=291

© **Women, Action, and the Media**

---

35. One response left this field blank.
36. This may be due to its relative size rather than relative harassment levels. For comparison purposes, Facebook's monthly active user count is more than four times that of Twitter. http://www.statista.com/statistics/272014/global-social-networks-ranked-by-number-of-users/

# 1.3

## TWITTER'S RESPONSE TO HARASSMENT REPORTS ESCALATED BY WAM!

After receiving a report of harassment, WAM! reviewers assessed the report and worked with the receiver of harassment to compile information needed to escalate the report to Twitter. WAM! used an online ticketing system to track conversations with reporters and receivers of harassment, as well as to group together multiple reports of the same case or individual reports associated with more than one case of alleged harassment. When WAM! reviewers were confident that the report could and should be escalated to Twitter, they used the company's standard abuse form to enter a report for special consideration. Upon receiving an escalated report, Twitter opened a ticket in their own system, following up up with WAM! reviewers with requests for more information and a notification of Twitter's final action.

One of WAM!'s goals in taking on the authorized reporter status was to offer a systematic evaluation of Twitter's responses to harassment reports—to develop a broad perspective grounded in multiple cases rather than the case-by-case experience of individual receivers of harassment. This section reports on Twitter's response to harassment reports escalated by WAM!.

### Table: WAM! Criteria for Evaluating and Escalating Reports

- Does the user demonstrate a pattern of targeting one or more other users with hate speech (misogynist, racist, homophobic, anti-trans, etc.) or violent speech?

- Is this an impostor account created to trick people, with the intention of undermining the reputation of the user the account is imitating?

- Has the user made threats of violence, or threats to release private personal information or photos?

- Has the user used Twitter to share someone else's private personal information, whether contact information, private and/or stolen photos, financial information, or other types of doxxing?

- Is the user using Twitter to spread verifiable lies about another user?

- Is the user encouraging others to harass someone, either online or off?

- Is the Twitter harassment just part of other harassment on other online platforms, or in offline life?

- Is one person being overwhelmed by harassment sent by a group of users?

## WHAT DID WAM! ESCALATE TO TWITTER?

WAM! reviewers used the following criteria for escalating reports. *(Part Two offers a more complete overview of the review process.)*

WAM! reviewers escalated 155 tickets, 43% of the317 incoming reports that they judged genuine:[37]

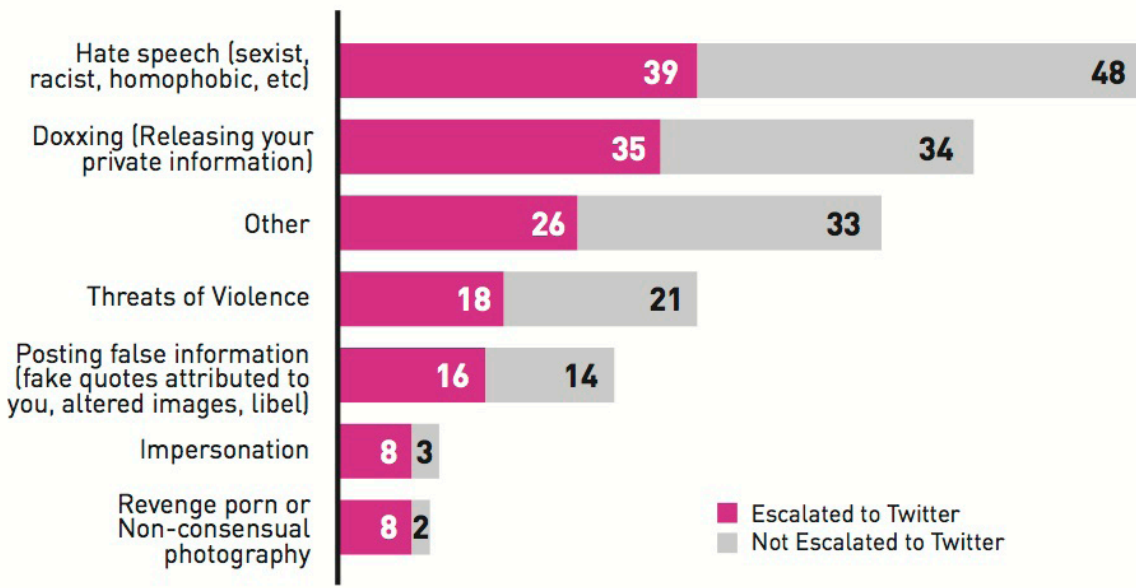

## WAM! Escalated Reports of Hate Speech, Doxing, Threats of Violence, and Other More Complex Cases

*Among reports submitted to WAM!, the number of reports escalated to Twitter, grouped by the self-identified harassment type from the WAM! reporting form.*

| Harassment type | Escalated to Twitter | Not Escalated to Twitter |
|---|---|---|
| Hate speech (sexist, racist, homophobic, etc) | 39 | 48 |
| Doxxing (Releasing your private information) | 35 | 34 |
| Other | 26 | 33 |
| Threats of Violence | 18 | 21 |
| Posting false information (fake quotes attributed to you, altered images, libel) | 16 | 14 |
| Impersonation | 8 | 3 |
| Revenge porn or Non-consensual photography | 8 | 2 |

■ Escalated to Twitter
■ Not Escalated to Twitter

SOURCE: WAM! Harassment Reports, Nov 6-26, 2014 n-317

**© Women, Action, and the Media**

37. Because WAM! used Twitter's own form rather than their own system to escalate tickets, we reconstructed the association with escalated tickets for our analysis. For more information on methods and limitations, see Appendix 1.3 and Appendix 1.4

### WHAT ACTIONS DID TWITTER TAKE?

The 155 tickets in WAM!'s internal system were associated 161 tickets opened by WAM! with Twitter. For each ticket in the Twitter system, a conversation occurred over email: Twitter notified WAM! that the ticket had been opened, asked for further information as necessary, and reported their final decision. These messages announced a variety of statuses, including *warning sent to account owner, account suspended, account previously suspended, request declined, information request, contacted user, issue resolved itself, authorization request, user not found,* and *user engaging with alleged abuser.*

Across these tickets, Twitter took action to suspend, warn, or delete the alleged harassing account in 55% of cases, reporting the following proportions of final decisions:

### WHAT KINDS OF HARASSMENT DID TWITTER TAKE ACTION ON?

By escalating an abuse report, WAM! reviewers indicated a belief that the alleged harassment constituted grounds for action by Twitter. Out of the 161 responses by Twitter linked with WAM! internal tags, the following chart shows the number of cases where Twitter took action— deleting, suspending, or warning accounts—

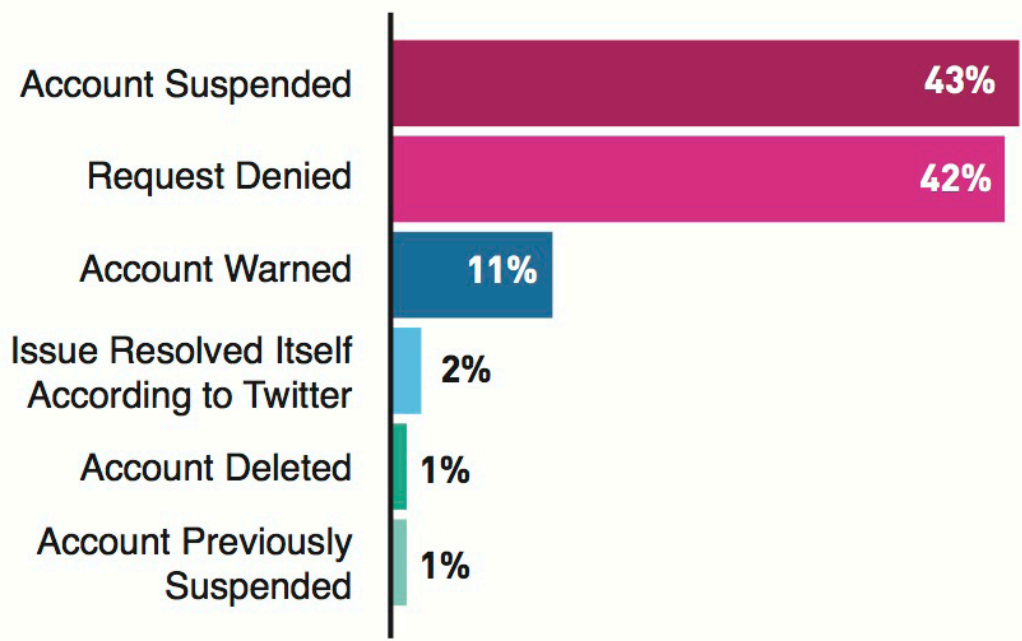

## Twitter Took Action to Suspend, Warn, or Delete the Reported Account In a Majority of Cases Escalated by WAM!

*Among reports submitted to WAM!, the number of reports escalated to Twitter, grouped by the self-identified harassment tpye from the WAM! reporting form*

| | |
|---|---|
| Account Suspended | 43% |
| Request Denied | 42% |
| Account Warned | 11% |
| Issue Resolved Itself According to Twitter | 2% |
| Account Deleted | 1% |
| Account Previously Suspended | 1% |

SOURCE: Tickets Opened escalated to Twitter by WAM!, from reports submitted to WAM! between Nov 6-26, 2014. n=161

© Women, Action, and the Media

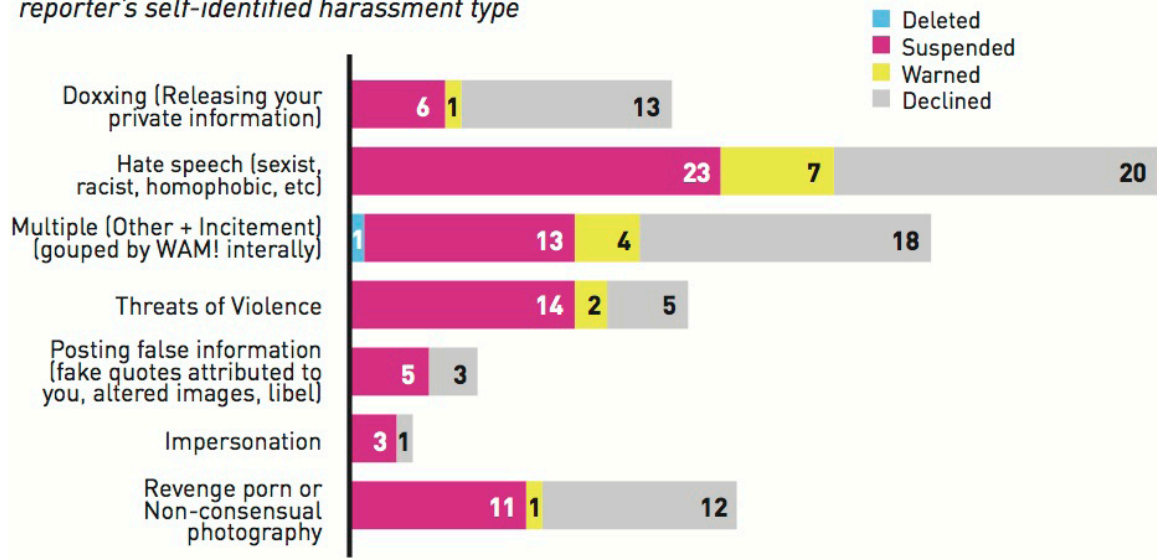

### Twitter's Actions Toward Allegedly Harassing Accounts In Response to Reports Escalated by WAM!

*Number of actions by Twitter to delete, suspend, warn, or decline action, in response to reports escalated by WAM!, grouped by the reporter's self-identified harassment type*

Legend:
- Deleted
- Suspended
- Warned
- Declined

| Harassment type | Suspended | Warned | Declined |
|---|---|---|---|
| Doxxing (Releasing your private information) | 6 | 1 | 13 |
| Hate speech (sexist, racist, homophobic, etc) | 23 | 7 | 20 |
| Multiple (Other + Incitement) (gouped by WAM! interally) | 13 | 4 | 18 |
| Threats of Violence | 14 | 2 | 5 |
| Posting false information (fake quotes attributed to you, altered images, libel) | 5 | | 3 |
| Impersonation | 3 | | 1 |
| Revenge porn or Non-consensual photography | 11 | 1 | 12 |

SOURCE: Composite data on WAM! Harassment Reports, Nov 6-26, 2014, linked with WAM! Ticketing system and coded emails of Twitter's responses to WAM!'s escalated tickets. In some cases, multiple harassment reports (and therefore multiple types of harassment) are sometimes associated with a single Twitter action.  n=161

© Women, Action, and the Media

compared to the number of cases where Twitter explicitly declined to take action, for each kind of harassment reported.

Overall, Twitter took action more often than they declined WAM! requests, with their rate of action varying by harassment type. In cases reported to involve threats of violence, Twitter took action to suspend or warn accounts in 16 cases, over three times more often than they declined (5). Where the initial reporter had selected 'Other' in order to indicate multiple types of harassment, Twitter took action at WAM!'s recommendation in 18 cases, declining in equal numbers. Twitter took action in more cases of hate speech (30) than they declined (20). Twitter declined to take action in most cases of doxxing, rejecting almost twice as many requests (13) as they took action on (7).

Both reporters of harassment and WAM! reviewers made assessments of the personal safety risk to receivers of harassment. All reports required this information; while not all WAM! tickets recorded WAM! reviewers' judgments of safety risk, many of them did.

■ Fears for personal safety were taken seriously by Twitter, which took action more often than not when reporters or WAM! reviewers expressed fear for the personal safety of the harassment receiver. However, Twitter declined to take action in 17 cases where the reporter feared for the

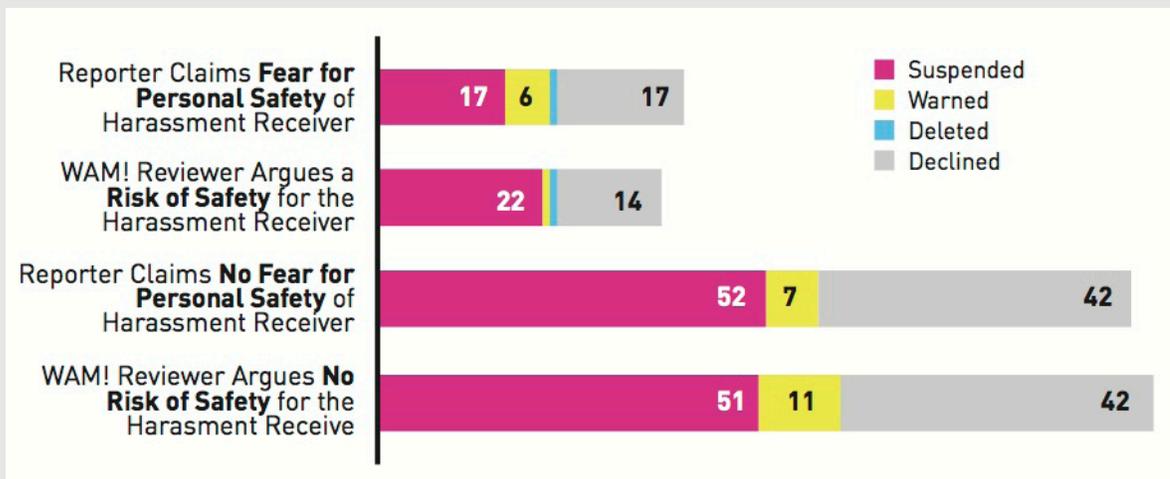

## Twitter Suspended Accounts Posing Safety Risk as Judged by Reporters and WAM! Reviewers

*Number of actions by Twitter to delete, suspend, warn, or decline action, in response to reports escalated by WAM!, grouped by reporter's or WAM! Reviewers' view on risk.*

**Reporter Claims Fear for Personal Safety of Harassment Receiver** — 17 | 6 | 17

**WAM! Reviewer Argues a Risk of Safety for the Harassment Receiver** — 22 | 14

**Reporter Claims No Fear for Personal Safety of Harassment Receiver** — 52 | 7 | 42

**WAM! Reviewer Argues No Risk of Safety for the Harasment Receive** — 51 | 11 | 42

Legend: Suspended / Warned / Deleted / Declined

SOURCE: Composite data on WAM! Harassment Reports, Nov 6-26, 2014, linked with WAM! Ticketing system and coded emails of Twitter's responses to WAM!'s escalated tickets. In some cases, multiple harassment reports (and therefore multiple types of harassment) are sometimes associated with a single Twitter action.  n=161

**© Women, Action, and the Media**

personal safety of the harassment receiver. Twitter also declined to take action in 14 cases where WAM! reviewers identified a personal safety concern.

■ **A lack of concern for personal safety from WAM! did not deter Twitter from taking action against an account**; Twitter frequently took action when neither the reporter nor WAM! had articulated fears for the personal safety of the harassment receiver.

### INTERPRETING DISAGREEMENTS BETWEEN WAM! AND TWITTER

Differences between WAM! and Twitter about the need for action do not necessarily reflect Twitter policy toward different kinds of harassment. For example, problems of evidence associated with certain kinds of harassment may have led Twitter to decline some kinds of requests more frequently than others. However, areas where Twitter declined large numbers of WAM! requests represent very real differences between the judgment of WAM!'s reviewers and Twitter's reviewers. Focusing on high risk areas of greatest difference can highlight aspects for improving support for people who are experiencing harassment.

**❗ Fears of personal safety:** In many cases, WAM! disagreed with the reporter's assessment of personal safety concerns. Furthermore, Twitter took action to suspend in more cases assessed by WAM! as safety risks than in cases assessed as safety risks by reporters. Differences between WAM! reviewers and reporters may highlight a need for better public information on assessing and dealing with personal risk associated with online harassment.[38]

**Doxxing:** The large number of doxxing reports declined by Twitter suggests that the practice needs to be examined more closely, especially given the high risk it represents for individuals.

**Multiple types of harassment:** harassment that was too complex to associate with a single preassigned category comprised one of the greatest challenges for WAM! reviewers, requiring considerable time and conversation. Such reports, which don't fit neatly into checkbox workflows, are arguably of very high importance: people who feel the most overwhelmed by harassment may be likely to use this option. While WAM! was successful at refining reports in ways that led to action by Twitter in half of these cases, Twitter declined to take action in the other half.

## PREDICTING TWITTER'S ACTIONS BASED ON TYPE OF HARASSMENT

Using the combined data from WAM!'s internal system and Twitter's responses, the following logistic regression model predicts the probability that Twitter will take action to delete, suspend, or warn an account based on the kind of harassment reported (accounting for WAM!'s assessment of the harassment receiver's personal safety risk, whatever judgment WAM! arrived at). In this model, the predicted outcomes for a given kind of harassment are compared to a 'reference category' of threats of violence—the category that had the highest probability of a response by Twitter. Positive estimates for each type of harassment indicate

### Predicting Twitter's Probability to Take Action toward an Alleged Harassing Account

*Logistic Regression predicting Twitter's action to delete, suspend, or warn an alleged harasser, from the type of harassment (with threats of violence as the reference category) and WAM!'s effort to assess the safety risk posed by the harassment in question.*

| | |
|---|---|
| WAM! Reviewers made note of evaluating fears of personal safety before escalating | **1.7091\*\*** **(0.563)** |
| Hate Speech (sexist, racist, homophobic, etc.) | -0.5436 (0.444) |
| Multiple (Other + Incitement) (grouped by WAM! internally) | -0.6665 (0.448) |
| Doxxing (Releasing your private information) | **-1.1445\*** **(0.552)** |
| Posting false information (fake quotes attributed to you, altered images, libel) | -0.7353 (0.543) |
| Impersonation | -0.6573 (0.814) |
| Revenge porn or Nonconsensual photography | -0.1405 (1.220) |
| Intercept | -0.4700 (0.403) |
| Pseudo R-squared | 0.0691 |
| Log-Likelihood | -97.621 |
| LLR p-value | 0.04310\* |

*Standard errors are reported in parentheses.*
*\* p<0.05, \*\* p<0.01, \*\*\* p<0.001*

SOURCE: Composite data om WAM! Harassment Reports, Nov 6 – 26, 2014, linked with WAM! Ticketing system and coded emails of Twitter's responses to WAM!s escalated tickets. In some cases, multiple harassment reports (and therefore multiple types of harassment) are sometimes associated with a single Twitter action. This subset of the overall 161 escalated tickets opened with Twitter only considers cases where Twitter took an action to delete, suspend, or warn an account, or explicitly declined action. n = 154.

© Women, Action, & the Media

an increased probability of Twitter taking action on average; negative estimates represent a lower probability of Twitter taking action on average.

On average, the odds ratio of Twitter taking action on a case of doxxing, compared to reported threats of violence, was 0.32; **the odds of action by Twitter were a third for doxxing as for threats of violence**. For example, on average, the probability of Twitter acting to delete, suspend, or warn in a case of doxxing (0.17) was twenty percentage points lower than the probability for threats of violence (0.38), where WAM! recorded an assessment of risk.

These results do not necessarily reflect a lower priority at Twitter towards doxxing. When the authors shared these results with WAM! reviewers, the reviewers suggested that many of the escalated reports of doxxing involved a 'tweet and delete' pattern. This tactic made collecting evidence that Twitter would deem acceptable difficult, since Twitter does not accept screenshots as evidence. As a result, WAM! reviewers believe that **Twitter's low probability of responding to doxxing is likely the result of cases where evidence has been removed by alleged harassers.**

On average, the odds ratio of Twitter taking action on reports escalated by WAM! that had received an assessment of personal risk versus ones that had not was 5.52; **the odds of action by Twitter were 5.5 times greater for tickets assessed for safety risk by WAM!**. For example, on average, the probability of Twitter acting to delete, suspend, or warn in a case of a threat of violence where WAM! had noted its assessment (0.78) was forty percentage points higher than cases where WAM! did not record its assessment (0.38).

Why was WAM!'s assessment of risk so important? A positive or negative assessment of personal risk was often an important part of the decision to escalate tickets to Twitter, although WAM! reviewers only added this assessment in some cases. These extra tags were often attached to cases where WAM! reviewers disagreed with the harassment reporter's assessment of risk, or where they were double-checking that assessment. The resulting escalated reports where WAM! reviewers invested that effort were more likely to receive action from Twitter than those where they did not.

In this model, other types of harassment—such as impersonation and revenge porn—were not associated with significant differences in the population of reports represented by this sample. This is not to say that Twitter did not take any action on these reports, or that they did not prioritize them, only that this model does not identify a significant relationship between these types of harassment and Twitter's probability of action.

## PREDICTING TWITTER'S ACTIONS BASED ON FOLLOWER COUNTS AND ACCOUNT AGE

Is Twitter more likely to warn, suspend, or delete newer accounts with smaller numbers of followers? Is Twitter more likely to take action on behalf of harassment receivers who have greater numbers of followers? To test theories of favoritism by Twitter in reports from and about different people, controlling for the kind of harassment, the authors fit a series of statistical models. This analysis **did not find any relationship between follower counts, account age, and Twitter's likelihood to take action.**[39]

---

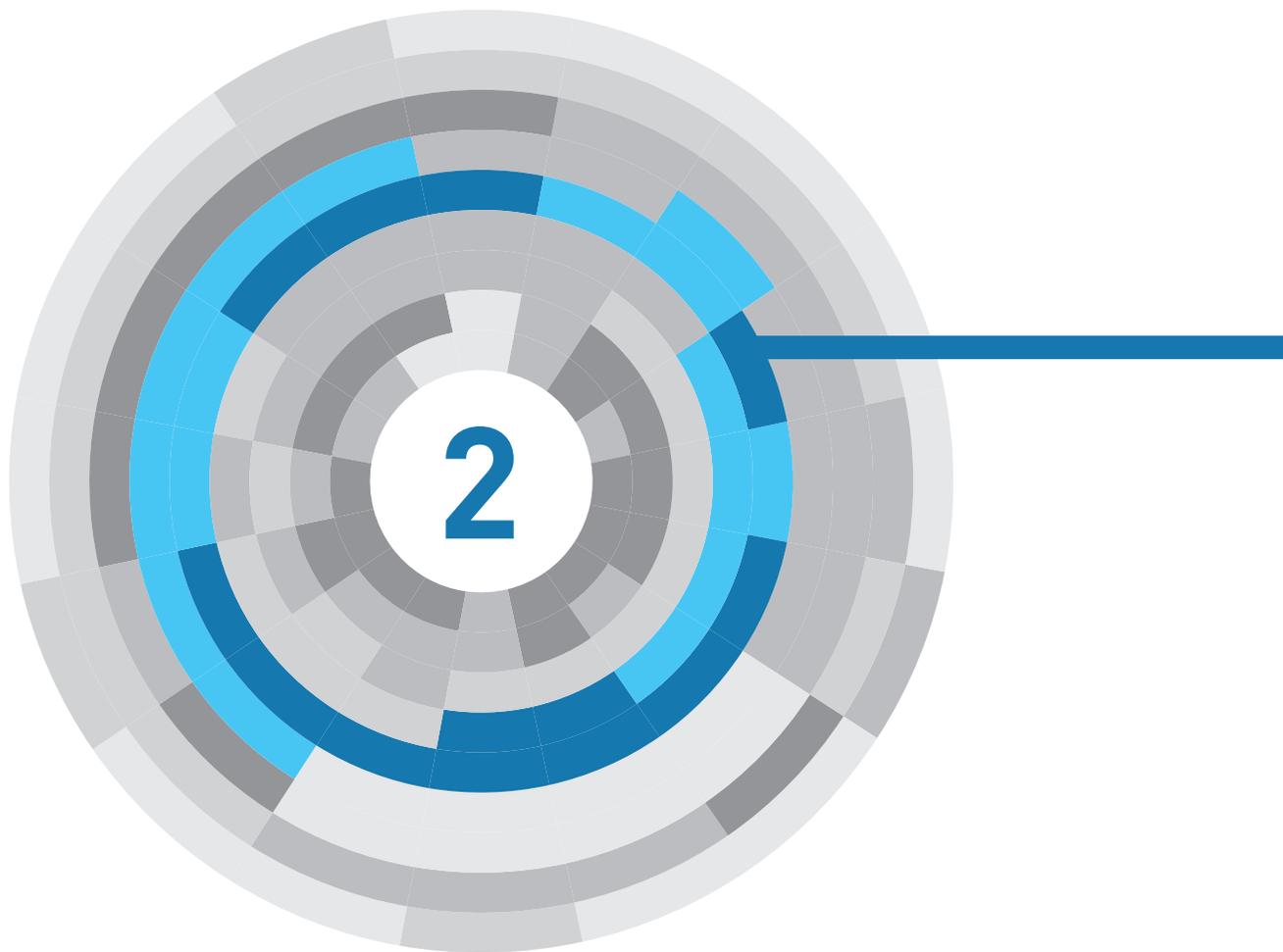

# PART TWO:
## REVIEWING HARASSMENT REPORTS

The WAM! project offers broader insights into the challenges of handling harassment reports. Teams that support Twitter and other social media companies carry out this largely invisible work twenty-four hours a day at large scales. Since few details of these content moderation processes are public knowledge, it can be difficult to assess the effort such work entails and the impact it has on reviewers, especially in aggregate.

For the WAM! project, the process starts with the authorized reporter relationship and the WAM! reporting tool. From there the process shifts to assessment and conversations with both reporters and the Twitter review team. Estimations based on WAM!'s volunteer experience can be considered a baseline of the effort involved: platforms are likely to have implemented support for harassment reporting that is at least as efficient as WAM!'s review process.

# 2.1

## THE AUTHORIZED REPORTER RELATIONSHIP

Typically when people report harassment on Twitter, they do so directly to the company, either via the reporting features on the Twitter platform or through a form on the Twitter website. In addition to these processes for individual reporting, Twitter has given a small number of organizations authorized reporter status. This enables these organizations to receive, review, and escalate reports to Twitter. Authorized reporters manage independent intake and assessment systems separate from Twitter's systems. Twitter's review team takes special note of reports arriving through these channels.

This project was possible because Twitter granted WAM! authorized reporter status. The people WAM! helped were helped thanks to that relationship. The data collected and the findings of this document are grounded in that relationship. WAM! is only one of Twitter's authorized reporters, and Twitter is not the only platform to make use of this kind of relationship. What are the consequences of the authorized reporter model? What does it achieve and what does it complicate?

### BENEFITS

For Twitter and other platforms, the authorized reporter relationship is a way to handle context and the challenges of determining harassment— more people with relevant expertise can review reports. For organizations like WAM!, it's an opportunity to serve as an advocate with powers that extend beyond platform reporting. In many cases, WAM! reviewers provided close personal attention and emotional support. WAM! also had the ability to direct individuals to resources outside the Twitter platform, such as a lawyer who had volunteered assistance.[40] Extending that advocacy, authorized reporters are also uniquely positioned to conduct and publish research like this report, with limited risks to a platform's staff or trade secrets.

### COSTS

The authorized reporter relationship also carries costs and risks. The relationship is unstable: The platform determines who is granted authorized reporter status, how the relationship works in practice, and whether or not it continues. Quality of support will vary with different authorized reporters. As a model, the authorized reporter relationship gives greater attention to groups that have associated advocates and experts, potentially at a cost to individuals reporting directly to Twitter. For authorized reporter organizations, the process requires substantial labor and can have mental health consequences for reviewers.

### A NEED FOR GREATER CLARITY

The authorized reporter relationship is not widely understood. Media articles about the WAM! project repeatedly mischaracterized the authorized reporter relationship. (A detailed analysis of media coverage of the WAM! project is included in Appendix 2.) Email exchanges between reporters and WAM! reviewers show that the boundaries between WAM! and Twitter were not always clear to reporters: Some reporters of harassment assumed WAM! had access to the Twitter system, specifically prior reports and correspondence or deleted/protected tweets. This **lack of clarity may cause people to misdirect reports of harassment and have unreasonable expectations of authorized reporter organizations, leading in turn to broken exchanges and abandoned reports.**

This lack of clarity extends to those *in* the authorized reporter relationship. The WAM!– Twitter correspondence shows some unevenness in how Twitter manages the authorized reporter relationship, particularly with regard to reports from people who were not the receivers of harassment (bystanders and delegates). Sometimes these reports were processed via WAM!'s authorized status. Sometimes WAM! received emails from Twitter requesting the name and @handle of the reporter, or requiring that the receiver of the harassment file a report. WAM! staff describe being unable to determine when either outcome would occur.

---

40. Specifically for serious cases where WAM! was otherwise unable to help.

# 2.2

## THE WAM!
## REPORTING TOOL

People who reported harassment to WAM! did so through an online form created and hosted by WAM!.[41] The design of the WAM! reporting form, accessible only through the WAM! website, shaped the data the project received and consequently the analysis contained in this document.

The WAM! reporting form asked reporters to categorize the type of harassment they were reporting, drawing on eight preassigned categories. The associated radio buttons of the form allowed reporters to select only one of these eight. This demonstrably influenced the categorization of harassment. Many reporters used the text box accompanying the final 'Other' category to indicate that the harassment they were reporting fell into a category not included in the preceding seven, such as spamming, stalking, inciting others to online harassment, and encouraging suicide. Many also used the 'other' category to indicate that the reported harassment fell into multiple categories. Reporters similarly pointed to multiple categories in the text box following the question, *Please describe in detail the harassment you are receiving*, as well as in later correspondence.

*An excerpt from the WAM! reporting tool; a complete version of the base form can be found in Appendix 3.*

....................................
41. In addition to the public-facing form hosted by WAM!, the reporting form was mirrored elsewhere in the event that the public site was attacked and had to be taken down. When the reporting period closed, the public-facing form was removed. The following analysis is based on the privately mirrored version of the reporting form. The complete base form is available in the appendix.

Notably, reporters may have also used the form in manners that departed from the intentions of the design. For example, though the WAM! reporting form did not explicitly allow anonymous reporting, individuals reporting on the behalf of others could supply incorrect names and email addresses—the WAM! team's declared process was to contact the receiver of the harassment rather than the reporter. (Similar practices were visible in the report trolling WAM! received.) Thus, although name and email were requested from all who reported on behalf of others, it is unknown how many these reporters used a genuinely identifying name and email address.

The attention and emotional state of reporters may have also affected which of the reporting form's directions reporters followed and how. The fourth question of the form asked reporters to *Enter Twitter handle being harassed (Do not include the @ symbol)*. This was paired with rollover text reading: *(Do not include the @ symbol)*. Despite this instruction, numerous reporters entered the @ symbol. There are multiple possible explanations for this: Reporters may have understood the @ symbol as integral to Twitter handles. Reporters may have been giving only cursory attention to instructions. It is worth highlighting, however, that **when reporting harassment, many reporters are likely to be in a high state of stress. Consequently, the ability to absorb directions and make decisions may be reduced.** This, in turn, can have both data and design implications.

## TWITTER'S REPORTING TOOLS

How is this similar or different from reporting to Twitter? Twitter updated its reporting tools soon after the WAM! project's reporting period officially closed. The main visible changes of these updates were the introduction of bystander reporting and a streamlining of the reporting process.[42] As this report was being finalized, Twitter began releasing images of further tool updates and has announced that it will be giving its review team new capabilities for responding to accounts that engage in harassment or abuse.[43] An update that allows reporters to choose to receive an email record of the harassment report to share with law enforcement has since followed.[44] As a brief overview, at the moment of writing, Twitter offers multiple ways to report tweets and accounts from mobile and web as well as support articles on online abuse,[45] on supporting others during such abuse,[46] and on trusted resources users can turn to for further help.[47]

Reporting tools take two primary forms: an in-platform tool and a web form. The current in-platform reporting tool[48] appears as a handful of consecutive screens, one question per screen. Except for the final screen, which offers a text box, it is composed entirely of required questions followed by predefined answers assigned to radio buttons. As with the WAM! form, these radio buttons allow only a single selection. The current web form[49] consists of 9 questions, 8 of which are required; answer formats include both radio buttons and text boxes. This form includes a text box for further detail, including types of harassment not covered by the form's predefined categories. It also requires different information of reporters than the in-platform tool, including a general location and an electronic signature.

# 2.3

## REVIEWING AND RESPONDING

Submitting the WAM! reporting form triggered WAM!'s review process, establishing a conversation with WAM!. This in turn often led to escalation—that is, WAM!, as an authorized reporter, submitted a report to Twitter on the individual reporter's behalf. Once reports were submitted to WAM!, they were automatically entered into the ticketing system used by WAM! reviewers to track the large number of cases reported (see *Figure 1*).

### HOW WAM! HANDLED INCOMING REPORTS

With the submission of the reporting form, the larger reporting process clicked into motion. Reports were routed to the internal WAM! ticketing system, an automated note was sent to the receiver or reporter of harassment as appropriate,[50] and WAM! reviewers began the work of reviewing incoming reports and starting a conversation via email with the receiver of harassment.

When a new report arrived in the system, a single WAM! reviewer assessed the report based on whether the report was genuine and included adequate information, following the links provided in the report and reviewing the Twitter timelines of the reported accounts. If the WAM! reviewer deemed the report genuine, the reporter (or receiver) would then receive a follow-up email from WAM!. If the report was deemed nongenuine, WAM! did not respond further.

.................................

50. As initially set up, the automated note that went to receivers of harassment included the contact information reporters of harassment had provided. This problem was brought to WAM!'s attention and addressed.

In the course of deciding which reports to escalate to Twitter, the reviewer asked the receiver of harassment (sometimes but not always the reporter) for additional information or documentation as necessary. Each report was reviewed along common characteristics assessing the nature of the harassment and the risk to the harassment receiver (see *Table: WAM! Criteria*).

### HOW MANY TICKETS DID WAM! IDENTIFY AS FAKE?

Out of all 594 incoming reports that were converted into tickets, WAM! volunteers judged 47% of them (277) to be fake. Most of these reports arrived in a single day (nearly 250), when WAM! received a high volume of requests from a bot. Note that WAM! reviewers reviewed reports without the administration privileges accessible to Twitter's team of reviewers, making determinations of falsehood more challenging.

### HOW LONG DID WAM! TAKE TO RESPOND TO REPORTS?

WAM! replied promptly to all tickets; across all 355 tickets where WAM! replied, WAM! responded within 386 minutes on average (6.4 hours), replying to 75% of reports within 10 hours. Every ticket was responded to in less than 24 hours.

### CONVERSATIONS BETWEEN WAM! REVIEWERS AND RECEIVERS OF HARASSMENT

This section, which combines quantitative analysis of conversation patterns and qualitative analysis of the kinds of exchanges that occurred, highlights three important types of exchange pattern: single contacts, broken exchanges, and extended exchanges. These terms do not imply inaction by WAM!; both 'broken' exchanges and extended exchanges were escalated to Twitter by reviewers.

Out of all 594 tickets, WAM! reviewers personally replied to 355 tickets after the initial automated response (60%), most often establishing a conversation with the person who was reportedly being harassed.

In the 285 cases in which WAM! did not send a personal response, most tickets did not receive any internal discussion from the WAM! team. Examples include duplicate submissions that were merged, or tickets that WAM! flagged as fake with no discussion. WAM! team members commented on average 3 times on the 17 tickets they discussed but did not reply to. Often this discussion was focused on assessment of the ticket as fake or not; tickets judged fake did not receive further communication.

In the shortest of these conversations, WAM! sent one follow-up message to the reported target of harassment. These 107 *single contacts* with no back-and-forth arose from the following situations:

- Reports immediately escalated by WAM! with no discussion, where Twitter took swift action in return, allowing WAM! to send a single response that Twitter had taken action

- Twitter having taken independent action to suspend the harassing account by the time WAM! looked at the incoming report

- Duplicate reports, often submitted by a second bystander

- Requests by WAM! for further information (especially in cases where the submitted evidence, such as screenshots, could not be accepted by Twitter) that were never returned

- Reports sent to WAM! where the target of harassment was handling the report through other channels, but wanted WAM! to incorporate their case into WAM!'s survey of harassment experience

- Reports focused on other social media platforms, such as Facebook

- Reports associated with cases where Twitter had previously denied WAM! requests

- Reporters asking for action that could not be escalated to Twitter

- 7 cases where WAM! promised to escalate the tickets to Twitter (including some duplicate reports from multiple sources),

promising to get back within 24 hours, and never replied to the target with the outcome;[51] all but two of these were eventually escalated to Twitter

- One case where WAM! received evidence in a language other than English and was unable to offer support

A *broken exchange* refers to an exchange that ends with an unmet expectation of response. The nonresponse may have occurred when it was the reporter's interaction turn or WAM!'s. Exchanges broken by reporters—that is, moments when reporters didn't respond—mainly followed requests from WAM! for more information, such as tweet URLs, previously assigned Twitter case numbers, or additional evidence.

Overall, exchanges broken by WAM! include all reports deemed fake and some triggering reports. After deciding a report was nongenuine (often decided without any email exchanges, but sometimes after limited exchange), WAM!'s team did not engage in further interaction. A small number of reports reports triggered intense emotional response in WAM!'s review team; a handful of these were accidentally left without response after reviewers disengaged to recover.[52]

Many conversations had substantial length. These *extended exchanges* included either a high number of emails or a high word count in a reporter's emails (for the most part WAM!'s team used brief email templates or wrote short personal emails). Extended exchanges were typically used by reporters for additional reports of harassment and emotional release.

Many times, reporters communicated additional instances of harassment and additional reports to Twitter within these follow-up

---

51. After further investigation, we determined that these gaps occurred after one of the three WAM! reviewers at the time stepped down after emotional trauma. In the transition, several tickets were missed.

52. Follow-up analysis shows that 5 out of 7 of these dropped tickets were escalated to Twitter, but that WAM! did not complete the conversation at the time.

email exchanges. Follow-up reports within the same WAM! ticket might include information about additional accounts or information on additional types of harassment. When additional accounts were mentioned in the conversation process, they were sometimes described as joint harassers engaged in campaigns of harassment. In some cases, additional accounts were reported for separate cases of harassment. Additional accounts were also reported in the context of the same user opening new accounts as reported ones were suspended. As an example of the variety of harassment types discussed within a single ticket, someone who initially reported the posting of unauthorized photos as part of a revenge porn attack might later in the email exchange provide details about having been doxxed as well.

Although the presence of many reported cases of harassment per WAM! ticket poses a challenge for the authors' quantitative analysis (which consequently undercounts the harassment reported to WAM!), **these conversations demonstrate the benefits of engaging in an extended conversation with people who are experiencing harassment, centering the process around the person rather than focusing it on the specific instance**.

These conversations also show evidence of emotional release, which took the form of venting (about the harassment or about Twitter's reporting process) and gratitude (repeated thanks for review/escalation of particular cases as well as for the WAM! project more broadly).

These extended exchanges show that for people experiencing harassment, the reporting process is much more than just harassment identification—it is part of coping emotionally with a traumatic experience. From the support perspective, the reporting process is an opportunity to establish trust and listen. **Processes optimized solely for stopping harassment are unlikely to address the larger impact of the harassment on the targeted user.**

## ASSESSING PERSONAL SAFETY CONCERN OF REPORTED RECEIVERS OF HARASSMENT

In each case, WAM! volunteers assessed the personal safety concern of receivers of harassment. Out of 317 reports not labeled fake, the reporter claimed a safety risk in 18%. In contrast, WAM! judged 25% of these reports to involve some kind of personal safety risk. In 30 cases, WAM disagreed with the reporter's assessment and concluded that there was no personal safety risk. In 54 cases, WAM! concluded that a safety risk did exist, even when the reporter didn't make that claim.

Qualitative analysis of reports where WAM! disagreed, concluding there was no personal safety risk, suggests that a limited number of reporters may have indicated fear of personal safety in order to emphasize seriousness and urgency rather than to indicate specific safety concerns. It is also possible that the project's focus on the Twitter platform meant that details that would explain physical safety concerns didn't surface in these reports: When following up, the WAM! team focused on acquiring information for assessment and escalation to Twitter. They didn't query reporters on why they felt a personal safety risk.

## WHAT DID WAM! ESCALATE TO TWITTER?

WAM! reviewers escalated 43% of reports they judged genuine. Why didn't WAM! reviewers escalate all tickets that they considered genuine? Aside from reports that reviewers did not judge to merit escalation due to the specifics of the interactions described, reports weren't escalated for the following reasons:

- some reports, especially bystander reports, were merged into a single ticket if WAM! had already escalated the issue

- some reports lacked the kind of evidence that Twitter required for escalation

- some reports were confusing, but when WAM! asked the reporter for clarifying information, the reporter did not respond

# 2.4

## THE WORK OF REVIEWING HARASSMENT

### WHAT CAN WAM!'S EXPERIENCE TELL US ABOUT THE WORK OF REVIEWING REPORTS?

Reviewing and responding to harassment reports can be challenging labor. It is work that carries considerable weight, urgency, and stress. Particularly with cases that might involve law enforcement, such as death threats or rape threats, a reviewer's decisions and speed of response can have profound effects on the safety of the receiver of harassment. Further, evidence of harassment is often emotionally difficult to read; reviewers may additionally encounter material designed by malicious reporters to cause harm to them.

At the same time, the labor of reviewers is largely invisible, with worker identities protected by companies and their work processes hidden to prevent harassers from exploiting loopholes. Consequently, with a few exceptions,[53] little is known about the nature of this work. The following, brief analysis of WAM!'s experience is offered to further support a more informed discussion about the work of reviewing harassment reports.

### REVIEWING HARASSMENT REPORTS: BY THE NUMBERS

While most of their work occurred during US daylight hours, the four WAM! reviewers responded to incoming email from reporters at almost all hours of the day. Reviewers replied promptly to every initial report deemed genuine: across the 355 tickets where WAM! reviewers engaged in conversation, their first response after the auto-response email was within 386 minutes of the initial report, on average replying to 75% of reports within 10 hours. All of these tickets were responded to in fewer than 24 hours.

Across the 21-day reporting period, WAM! reviewers:

- Assessed 640 incoming reports (30/day):

  assessed genuineness of each report (assessing up to 255 in one day)

- checked evidence in each report

- Sent a total of 1226 messages in the ticketing system (58 per day)

- Wrote 59,243 words in the ticketing system

- Examined more than 531 unique allegedly harassing accounts and 179 unique receivers of harassment (34 per day)

- Responded personally to 355 reports (17 per day)

- Evaluated or added 3,628 tags in the internal ticketing system (173 per day)

- Escalated 155 tickets to Twitter (7 per day)

- Carried out at least 186 exchanges with Twitter, as measured through Twitter's responses[54] (9 per day)

### MENTAL HEALTH COSTS OF REVIEWING HARASSMENT REPORTS

Reviewing and responding to harassment reports took a toll on the WAM! team. Reviewers reported secondary trauma from reading abusive and threatening tweets and the backstories of reporters. Symptoms included anxiety, sleeplessness, loss of concentration, depression, the triggering of past PTSD, and irritability, among others.

---

53. See the Introduction for discussion and references on the work of reviewing harassment reports

54. The actual number is higher, but the authors don't have access to this data

Some members of the WAM! staff and board themselves received harassment on Twitter as a result of the reporting project. As the names of the people on the reviewing team were never publicly released, this was regardless of whether those individuals were part of that team or not. This harassment included hate speech, distribution of photoshopped images and false information, and rape and death threats.

# 2.5

## THE PROBLEM OF EVIDENCE

Evidence presents a serious challenge both for people reporting harassment and for people reviewing reports. As data across the WAM! project demonstrates, this challenge is often exacerbated by assumptions about the forms harassment will take.[55] When reporting tools and review processes accept only certain modes or formats of evidence, it becomes difficult to submit evidence for forms of harassment that fall outside these assumptions. Platform responses such as suspension and deletion in turn affect the evidence available for reporting harassment to law enforcement and other channels.

### TWITTER & EVIDENCE

Twitter requires reporters to provide URLs for examples of harassment. This assumes that tweets are the mode through which harassment is being carried out. In the event that harassment occurs in some other way—for example, through exposure to violent or pornographic profile images or usernames via follower/favorite notifications—reporting harassment becomes complicated. Such harassment currently cannot be reported using Twitter's in-platform tool. While it can be reported via Twitter's web form, the reporter is still also required to provide a URL.[56]

Initially, WAM! encouraged reporters to submit screenshots of harassment. Only later did the WAM! team realize that **Twitter does not currently accept screenshots as evidence.** While the URL of a tweet is self-authenticating, an image presented as a screenshot could be the product of digital manipulation. For Twitter reviewers, assessing a screenshot also requires a greater expenditure of resources: the reviewer must locate the harassing element within the screenshot and then match it to a relevant object—account, tweet, image, etc.—within the Twitter system.

**Screenshots, however, are currently the primary means for capturing harassment that uses the 'tweet and delete' tactic—a tactic that cannot easily be captured via URLs.**

A harassment tactic seen on many platforms is for authors to delete harassing messages after the messages have been responded to/seen. This removes a critical piece of context that affects the way later viewers—and reviewers—of an exchange assess aspects like validity, emotional tone, and reasonableness of response.

On Twitter, when a harasser deletes a harassing tweet it not only affects later assessment of exchanges, it eliminates the URL associated with the tweet. Which means that **if the tweet wasn't reported immediately, pre-deletion, reporting the tweet via URL is now impossible**.[57] Further, Twitter's data retention policies[58] suggest that **even if a URL *is* reported in time, if the associated tweet is deleted before being seen by reviewers, the URL may not continue to function internally—that is, the evidence may be inaccessible even to Twitter reviewers**.

---

55.  This is visible, for example, in the richly detailed personal accounts in WAM!–reporter correspondence, but also in WAM!–Twitter correspondence; further, WAM! staff attribute Twitter's response rate to doxxing to evidence complications related to 'tweet and delete' patterns.

56.  On the web form, in conjunction with a text box labeled 'Further description of problem' Twitter requests reporters detail harassment that doesn't fit the tweet category. That web form can be found here: https://support.twitter.com/forms/abusiveuser.

57.  There are many reasons a harassing tweet might not be immediately reported; for example, emotional well-being may require distance before dealing with harassing or abusive messages.

58.  https://support.twitter.com/articles/41949-guidelines-for-law-

Taking a screenshot creates a digital record that has a permanency outside the Twitter system, a record outside the control of the alleged harasser. A screenshot can be shared even after a tweet has been deleted or a following or favoriting action has been undone. Further, exchanges in the WAM!–reporter correspondence suggest that many people find taking a screenshot easier and more familiar than locating the URL of a tweet and saving it. **Harassment is thus likely recorded earlier and more often with screenshots**.

## LAW ENFORCEMENT & EVIDENCE

Account suspension can slow or limit a harasser's ability to continue harassment on a single platform, particularly if suspension is linked to an IP address or telephone number[59] and hinders the creation of fresh accounts. However, when accounts are suspended, harassing or abusive tweets are no longer visible. The target of the harassment thus loses the ability to show the harassment or abuse directly to law enforcement. **The same complications created by the 'tweet and delete' tactic are also consequences of the process of 'review and suspend.'** Twitter itself advises users who contact law enforcement about threats to "document the violent or abusive messages with print-outs or screenshots."[60] This pattern and its consequent problems, however, are not unique to the Twitter platform.

As this report was finalized for publication, Twitter announced that reporters of harassment will now have the option to receive a record to share with law enforcement. This record summarizes the harassment report, but doesn't include evaluation or response.[61] Full assessment of this new option is beyond the scope of the current analysis, but it suggests that some of the difficulties previously experienced in conveying the seriousness of online harassment to law enforcement may now improve, at least with regard to the Twitter platform.

Even with this new option, when an account is suspended for harassment, no explanation is provided on the Twitter platform itself. All that users—including law enforcement—see at the URL of a suspended account is a blanket suspension notice. As a result, **suspensions for harassment or abuse are indistinguishable from suspensions for spam, trademark infringement, etc.** This contrasts with Twitter's policy for explicitly marking 'withheld content'— tweets or accounts censored in particular locations due to government requests.

For those reporting harassment to law enforcement, an indication on the Twitter platform of suspension due to harassment would be helpful. Further, such public acknowledgment could potentially reduce experiences of isolation or stigma, deter harassing behavior, and provide more robust data for analysis.

# ACKNOWLEDGMENTS

This report and WAM!'s overall project has been made made technically possible through the remarkable, substantial work of Adria Richards, who designed and set up WAM!'s reporting system, advised on matters of data, and offered critical advice on coordinating WAM! volunteers to code emails Twitter's responses. The authors share the deepest respect for Richards, who has responded to the damage of online harassment on multiple platforms and marginalization in the media with persistent creativity, resourcefulness, and care for the needs of others.

WAM! director Jamia Wilson offered ongoing coordination and collaboration to support this report. Willow Brugh offered helpful copy-editing feedback. The unsung heroes of this work include the unnamed WAM! reviewers and the WAM! volunteers who manually coded nearly two hundred messages from Twitter to support our analysis of Twitter's responses. We are also grateful to everyone who trusted WAM! with their reports, including false-flaggers, for the contribution they have made to our shared understanding of online harassment.

The peer review process was managed by Zeynep Tufekci, Assistant Professor at the University of North Carolina Chapel Hill, who generously facilitated a double-blind peer review for this paper. Over a period of 6 weeks, Tufekci worked with us to design the review process, recruited anonymous reviewers, sent them copies of our draft, compiled reviews from many respondents, gave us guidance on how to respond to the reviews, compiled our resubmissions, compiled final reviews, and confirmed the final response of reviewers. The authors are grateful for the detailed feedback and high standards of our reviewers, which led to a much clearer, stronger report.

Media analysis was made possible by the MediaCloud project at Harvard and MIT, who while not officially sponsoring this research, granted us access to their media archive and analysis tools. Ethan Zuckerman, director of the MIT Center for Civic Media, offered early, valuable advice on the approach to take with this report.

# APPENDIX 1: DATA, METHODS & LIMITATIONS

This report makes use of full data from the WAM! authorized reporting period of November 6 to December 27, 2014, including 811 submissions via the WAM! Twitter Reporting Tool, 640 tickets within WAM!'s ticketing system for handling incoming reports, and 185 messages received from Twitter in association with cases escalated by WAM! to Twitter. The authors were given access to WAM!'s ticketing system to observe harassment reports in the context that reviewers handled them, as well as internal documentation. We also held an ongoing, collaborative conversation with one of the WAM! reviewers for more information on WAM!'s process and context of our observations.

## 1.1 METHODS: SUPPORT PROFILES, CONVERSATION PATTERNS, REPORTING TOOLS

This analysis draws on the 811 initial reports, the 640 tickets and associated threads of WAM!–reporter correspondence present in the WAM! ticketing system, and the 185 messages of WAM!–Twitter correspondence organized via Grexit tagging, as well as internal WAM! documentation, personal accounts of WAM! staff and volunteers, and experimentation with the WAM! reporting tool and Twitter reporting tools.

We approached WAM!–reporter correspondence and WAM!–Twitter correspondence using the iterative process of grounded theory to identify salient themes. We accessed WAM!–reporter correspondence via WAM!'s ticketing system (built with ZenDesk) to emulate the context reviewers experienced when reviewing and responding to harassment reports; we similarly accessed WAM!–Twitter correspondence via gmail and Grexit tagging (see 1.4 below for more information on the Grexit tagging process). We selected cases for additional discourse analysis based on three strategies: keywords arising from discussion with WAM! staff, preliminary quantitative findings, and chain referral based on report contents. Support profiles constructed from these methods are offered as composites in order to protect reporters' privacy while highlighting important themes in the dataset. Observations on conversation patterns in section 2.2 are offered without quotes for similar reasons.

Internal documentation included the set of response templates WAM! created to interact with reporters via its ticketing system (only some of which were used in practice) and criteria for assessing reports, explicitly articulated by WAM! on November 18, 2014. Personal accounts of the WAM project were gathered via semi-structured group discussion paired with targeted followup questions via email, from Jaclyn Friedman, WAM!'s executive director during the reporting phase of the project and the project's primary reviewer, and Adria Richards, volunteer designer of the WAM Twitter Reporting Tool and the WAM! ZenDesk ticketing system. Internal documents and personal accounts provided contextualization and were used nonexclusively to prompt further in-depth qualitative assessment. In a few cases, members of the WAM! team provided specific explanations for this report's findings based on their personal experiences with the project; those insights are so noted. We assessed reporting tools through comparative design review and direct reporting to Twitter by one of the authors as a nonauthorized reporter.

## QUALITATIVE ANALYSIS DRAWING ON THESE DATA AND METHODS IS WOVEN THROUGHOUT THIS DOCUMENT. 1.2 METHODS: TYPES OF HARASSMENT

To consider what kinds of harassment were reported and how people used the reporting form, we carried out analyses of individual reports alongside aggregate summary statistics on the overall patterns of reporting. Next to the qualitative analysis described above, quantitative analysis of types of harassment is conducted on the subset of 317 incoming harassment reports validated as genuine by WAM! reviewers.

To generate this list of validated reports, we matched the 811 incoming harassment reports with the 594 reports that were associated with records in WAM!s internal ticketing system.[62] The internal WAM! ticketing system, ZenDesk, supports tagging features that were applied by WAM! reviewers to indicate their individual judgment on a ticket and its associated conversations. Claims about WAM! reviewers' judgments, such as judgments about the genuineness of a report, are based on these tags.

Since incoming reports and ZenDesk tickets did not share a common unique identifier, we matched these records by extracting structured data from ticket text that corresponded to fields in incoming reports. These matches were verified by human coders. Fields not matched by this automated technique were matched by a manual coding process that followed the same coding rules as the automated verification.

The questions on the WAM! Twitter Reporting Tool are listed in full in Appendix 3.

Several hundred incoming reports of harassment were discarded as nongenuine throughout the process. From November 6 to December 27 2014, WAM! received 811 incoming reports. Of these, 31% of reports (255) arrived on the 11th of November during a gap in WAM!'s spam filtering. Of all reports, 594 (73%) were turned into tickets within WAM!s internal processing system. Of these reports, 277 (47%) were flagged as fake reports by WAM!. Although qualitative analysis shows that a small number of these reports may in fact have been in good faith, we use the subset of 317 non-fake reports in our analysis of incoming reports.

## 1.3 METHODS: ANALYSIS OF WAM!'S REVIEWING ACTIVITY

In the analysis of the activity of reviewing and responding to harassment reports, we consider all 640 WAM! tickets created between the 6th of November and the 26th of November, not just ones labeled as genuine, and not just the ones associated with harassment reports.[63] We also incorporate conversations with one of the WAM! reviewers, as well as records made by WAM! on the November 18 that documented their internal process for reviewers.

## 1.4 METHODS: TWITTER'S RESPONSES

To consider Twitter's actions on tickets escalated by WAM! we used the collaborative email tagging system Grexit to coordinate a group of WAM! volunteers to manually code all 185 messages from Twitter, associating the official Twitter *abuse report* ticket number, the date of the message, the Twitter account being reported, and the set of WAM!'s internal tickets associated with the alleged harassing account in the Twitter abuse report, leaving out cases reported to WAM! after the moment of escalation. Volunteers also coded Twitter's responses.[64] The resulting linked dataset supports our analysis of Twitter's actions in response to 161 abuse reports open by WAM! with Twitter. Note that multiple reports to WAM! sometimes fed into a single ticket in WAM!'s system, and a single WAM! ticket sometimes resulted in several tickets being filed with Twitter.

---

62. The remaining incoming reports were empty or duplicates. For example, 255 duplicate reports arrived on November 11 during a gap in WAM!'s Captcha filtering; only some of these were converted into tickets in WAM!'s internal system. Three days earlier, on the November 8, WAM! was warned anonymously via its reporting form that 'GamerGate' was planning a campaign of fake reporting and spam, aiming to submit 10,000 spam reports over the subsequent few days. This was a significant overstatement compared to the number received. Beyond the reporting timeframe, we counted 143 empty reports and 37 additional reports that were never added to the WAM! ticketing system, arriving after the conclusion of the reporting period.

63. In this section, unlike the analysis of form submissions, we consider all tickets, including ones later labeled as fake by WAM! It is important to note that some of the 640 tickets were prompted by something other than a report. Only 594 reports in total were associated with tickets.

64. Twitter's responses included: account deleted, warning sent to account owner, account suspended, account previously

We also collected information on the number of followers of involved accounts by querying the Twitter API on December 29, 2014 for all 9,755 accounts mentioned in any report. Accounts that were suspended, deleted, nonexistent, or private accounts could not be queried, including allegedly harassing accounts that were permanently suspended by Twitter, as well as receivers of harassment who deleted their accounts or made them private before our sample.

In addition to the model reported in section 1.3, we  attempted to fit models for actions by Twitter where account information was available for the allegedly harassing accounts, controlling for harassment type and WAM! flags (n=61). We also attempted to fit models for actions by Twitter where account information was available on the receiver of harassment (n=122). No significant terms were found in any of these models. Finally, we attempted to fit models for actions by Twitter where information was available about the followers of allegedly harassing accounts and also followers of receivers of harassment (n=57). This final set of logistic models failed to converge despite using several estimation methods and model selection approaches.

Across these models, we failed to reject the null hypothesis that there is no relationship between follower counts of alleged harassers or follower counts of receivers of harassment and Twitter's likelihood to take action. However, our ability to answer this question was limited by our lack of access to information on accounts that were suspended or deleted by Twitter before the date we sampled follower information (resulting in small and biased sample sizes and a loss of statistical power).

## 1.5 METHODS: GAMERGATE ANALYSIS

In this analysis, we assess the proportion of reporting activity associated with GamerGate. We used the most inclusive available dataset of potentially GamerGate-related accounts with data from *ggautoblocker* by Randi Harper,[65] featuring 9844 accounts acquired from its public block list on the *BlockTogether* system on December 29, 2014. We chose *ggautoblocker* due to its extremely inclusive definition of potential GamerGate related accounts—the system auto-blocks anyone who follows a certain number of identified GamerGate participants. Although this inclusive blocking practice has been criticized for over-blocking accounts,[66] the inclusivity of this dataset is appealing for our uses, to search as inclusively as possible for any potential GamerGate connection.[67]

## 1.6 METHODS: MEDIA COVERAGE

To measure media impact, we used Media Cloud tools to analyze online media coverage of WAM!'s authorized reporter status with Twitter, in the context of coverage of harassment and abuse on Twitter more generally. Media Cloud is a suite of tools developed by the Berkman Center for Internet and Society at Harvard University and the MIT Center for Civic Media and made available to scholars researching online media. These tools monitor online content for over 50,000 sources worldwide on an hourly basis, downloading an average of 500,000 stories per day. In this analysis, we consider the

---

suspended, request declined, information request, contacted user, issue resolved itself, authorization request, user not found, user engaging with alleged abuser

65. See  http://blog.randi.io/good-game-auto-blocker/

66.  Wofford, Taylor. One Woman's New Tool to Stop Gamergate Harassment. Newsweek, Nov 29, 2014.

67.  Since there is no rigorously developed list of GamerGate targets, we limit this analysis to alleged harassing accounts

media during and after the WAM! Twitter Project in the wider context of media from 1 January 2014 to 15 February 2015.

## 1.7 LIMITATIONS: REPRESENTATIVENESS OF FINDINGS

How representative is WAM!'s data of people submitting reports to Twitter?

## 1.7A LIMITATIONS: VOLUME, GENDER, LANGUAGE & GENUINENESS

The reporting project was a WAM! initiative. People learned about the project—and how to participate—through information spread by news media and social media. Although Twitter established WAM! as an authorized reporter, the project wasn't integrated into Twitter's reporting tools. Thus, in order to report abuse or harassment to WAM!, individuals had to go directly to the WAM! web form. This point of entry undoubtedly influenced the data received. Three immediate aspects likely to have been impacted are volume, focal point of harassment, and language. The media analysis conducted offers information regarding who might have learned about WAM!'s form. Since the WAM! reporting form did not request any information about traditional sociological variables associated with identity, such as gender, race, age, class, sexual preference, etc, this analysis cannot report on those variables.

**We can expect WAM!'s report volume to be lower than Twitter's,** since reporting required knowledge of the project and navigation outside the Twitter platform.

**We can expect WAM!'s data to be more focused on gender-based harassment and abuse** than Twitter typically receives. WAM! is a nonprofit dedicated to issues of gender justice in the media and the announcement of WAM! Twitter project specifically solicited gender-based harassment and abuse**.** Furthermore, during the escalation process, the WAM! team declined a number of reports as outside the project scope as the reports didn't show evidence of "gendered harassment."

**The project was announced in English and its reporting tool was in English.** Though the project received exposure in regions where other languages are in use, the entire WAM! dataset is English-focused.[68]

The WAM! team had no explicit process for determining if reports were nongenuine; rather, they drew on personal communicative competence and team expertise. Non-systematic spotchecking suggests **a limited number of determinations of nongenuineness may have been in error**.

## 1.7B LIMITATIONS: WAM! IS NOT TWITTER

The consequences of reporting to WAM! differ from those of reporting to Twitter. This influences the types of reports received. Three types of reports found in the WAM! dataset that may vary from those likely received by Twitter include *resubmission of older reports*, *reports by media representatives*, and *trolling*.

The dataset includes numerous reports and mentions of **older examples of harassment and abuse that had previously been submitted to Twitter** without—in the reporters' estimations—result. Many were submitted in hopes that WAM! could effect some sort of change. In addition, some may have been submitted with the intent of having their case included in WAM!'s analysis. While Twitter

......................................
68.  The project only received one report in a language other than English, and was unable to address it due to lack of the necessary language abilities.

undoubtedly receives follow-up communications and duplicate reports, this particular form of revisiting older reports is likely specific to the WAM! dataset and WAM!'s role as an external agent.

The dataset also includes reports from **individuals who appear to be associated with news media outlets.** Many journalists receive harassment or abuse online. While the reports WAM! received appear to be genuine in this regard, it is possible that some of those submitting the reports were also doing so from a professional desire to learn more about the WAM! project and its outcomes. Twitter undoubtedly receives reports driven by similar motivations; how similar WAM!'s version of these reports is to Twitter's is unknown.

The project also received **trolling, spam, and other communications.**[69] Trolling involved the intentional submission of nongenuine reports. WAM!'s external role and substantial media coverage likely affected perceptions of the consequences of submitting a nongenuine report. Twitter undoubtedly receives trolling, spam, and similar nongenuine reports, but the WAM! experience is likely different from Twitter's.

## 1.7C LIMITATIONS: WHO DIDN'T REPORT TO WAM!?

It is impossible to measure accurately who didn't report to WAM! and why. That said, four general categories can be pointed to: People who didn't know about the project, people who felt the abuse/harassment they encountered didn't fall within WAM!'s project description, people disinclined toward external reporting, and people who chose to respond to abuse or harassment in ways other than reporting.

## 1.7D LIMITATIONS: HARASSMENT CASES PER TICKET

As noted in the section entitled Conversations Between WAM! Reviewers and Receivers of Harassment in the text, the presence of multiple cases of harassment per WAM! ticket—often shared via extended conversation or after reporters indicated the 'other' category on the WAM Twitter Reporting Tool—poses a challenge for quantitative analysis, as using the data from the set of initial reports as a harassment count undercounts the total instances of harassment reported to WAM!

---

69. How spam and communications other than reports of harassment or abuse on Twitter—e.g., general contentions that the WAM! project itself was a form of harassment—were handled in the dataset is documented in the methods section.

# APPENDIX 2: MEDIA COVERAGE

## MEDIA COVERAGE OF THE WAM! TWITTER REPORTING PROJECT

One of WAM!'s goals was to raise awareness about the issue of online harassment. WAM!'s authorized reporter status with Twitter was widely covered by media in multiple countries and began the widest covered story about online harassment on Twitter in 2014, offering a largely positive reflection on both organizations in a year of controversies about harassment online.[70]

## MEDIA COVERAGE OF THE WAM! TWITTER PROJECT

We queried the entire Media Cloud dataset for the period between 1 November 2014 and 15 February 2015, and found 204 stories that mention WAM! by name when talking about WAM!'s authorized reporter status with Twitter. The majority of these (146 stories, or 71.5%) ran between 6 November and 14 November, when the harassment reporting pilot project was first announced. There was a moderate second spike in mentions around 5 February 2015, after the *Verge* published internal memos from Dick Costolo about how Twitter handles abuse. There were also two additional, somewhat smaller spikes in mentions of WAM!'s role—the first on 2 December, when Twitter announced changes in its abuse reporting tool, and the second around 8 December, during the controversy surrounding the spread of personal information about the UVA rape victim profiled in *Rolling Stone*.

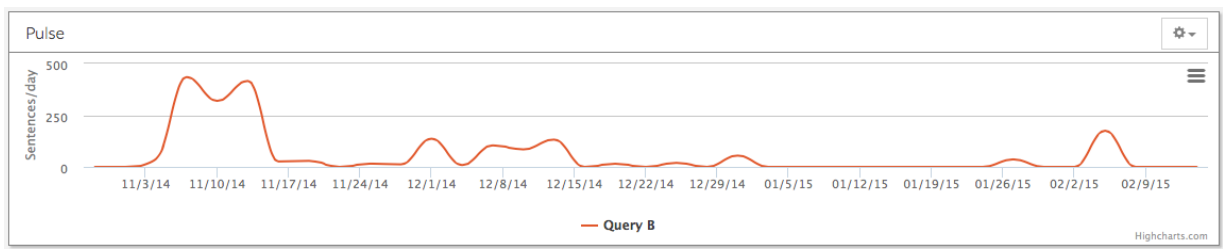

***Sentences per day:*** *Timeline of 204 stories mentioning WAM! when talking about the harassment reporting project, expressed in sentences per day (includes all sentences from each story)*

## WHO COVERED THE WAM! TWITTER PROJECT

Coverage of WAM!'s authorized reporting status received the most coverage in online-first media outlets, tech media, and US media. Our 204 story sample also includes media from 21 different countries and two international outlets not associated with any particular country.

---

70. As measured by the number of sentences per day in the Media Cloud database that specifically mention harassment or abuse on Twitter.

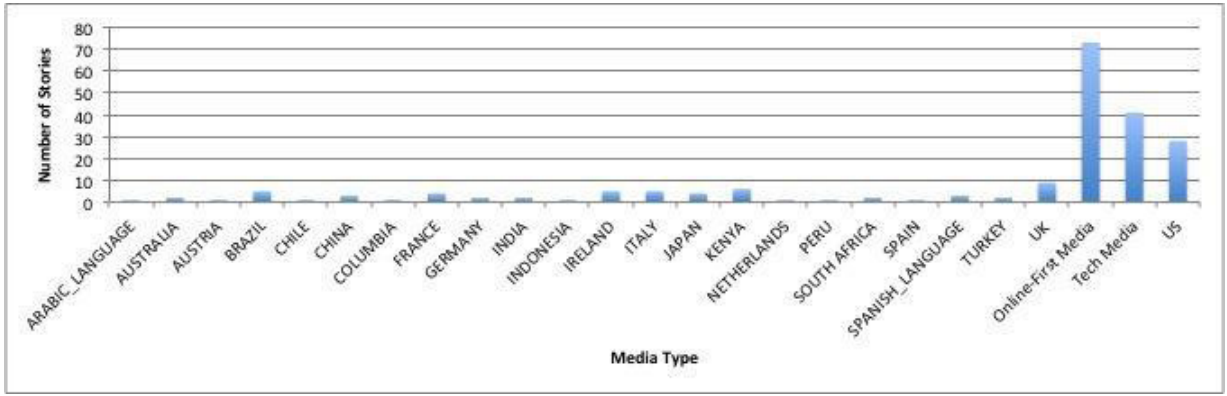

*Geographic distribution of 204 stories from the Media Cloud database mentioning WAM! when talking about the harassment reporting project. Many of the online-first and tech media are not associated with any particular country* [71]

## TONE OF MEDIA COVERAGE ON THE WAM! TWITTER PROJECT

**The 172 English-language stories that mention WAM! predominantly view the project in a positive ligh**t.[72] Most stories about the project itself state that harassment and/or abuse—especially of women—is a significant and well-known problem on Twitter, and that the harassment reporting project is both warranted and a welcome step. Stories that are about something other than the harassment reporting project overwhelmingly reference the WAM! project as a positive step, especially when the story's main focus is criticism of another platform (e.g., Facebook, Kinja, Uber) or the harassment and abuse of women online in general. The most common critique of the project is from authors who are sympathetic to its aims but feel Twitter should be doing more. A very small minority (six stories, or 3.5%) decry the harassment reporting project as "censorship" (by either "feminists" or "SJWs"[73]); notably, three of these six stories are written by the same blogger. Overall, these 172 stories represent the most significant positive coverage of Twitter with respect to harassment and/or abuse during the 13-month period studied (see elaboration below).

The table below shows the 100 words used most frequently in the English-language subset.

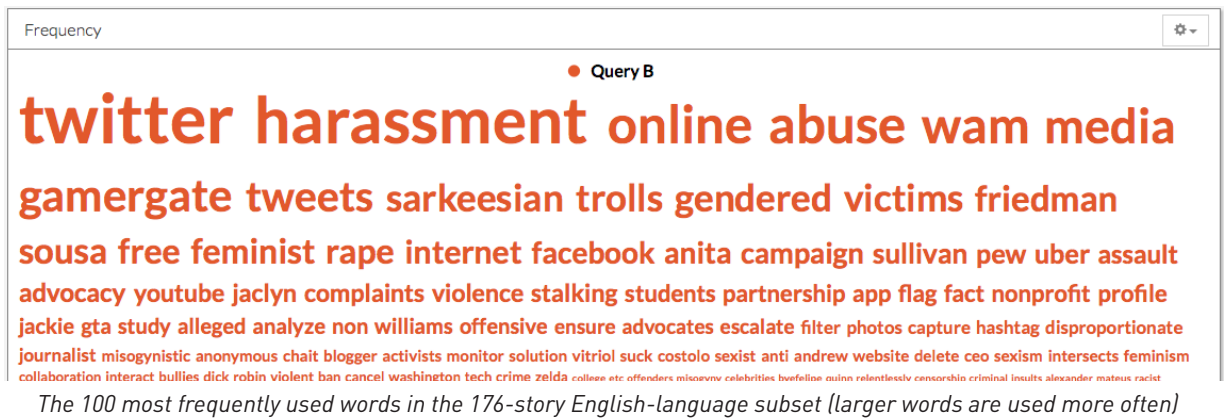

*The 100 most frequently used words in the 176-story English-language subset (larger words are used more often)*

## MISREPRESENTATION OF WAM!'S AUTHORIZED REPORTER STATUS

Despite the efforts of both WAM! and Twitter to explain that this project was not a collaboration between the two organizations, journalists did not accurately represent WAM!'s authorized reporter status. While media coverage initially characterized the project as either a "collaboration" or, more frequently, a "partnership," by December 2014 journalists consistently described the project as a "partnership."

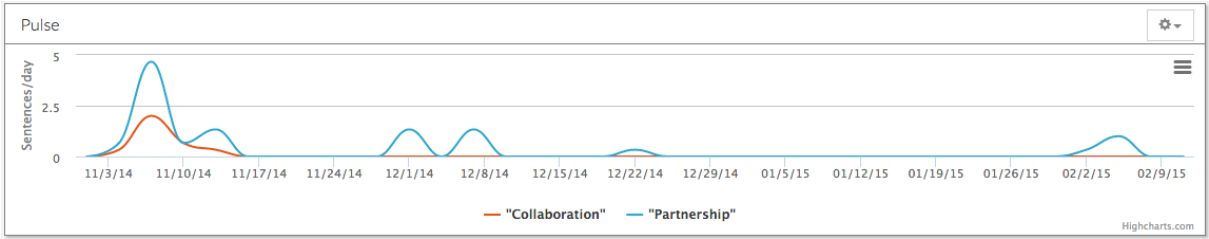

*Sentences about the project that included a "collaboration" term compared to sentences that included a "partnership" term.*

The term that more accurately describes the relationship—"authorized reporter status"—rarely made an appearance in media coverage of the reporting project. The below graph compares "collaboration" or "partnership" characterizations against the word "authorized."

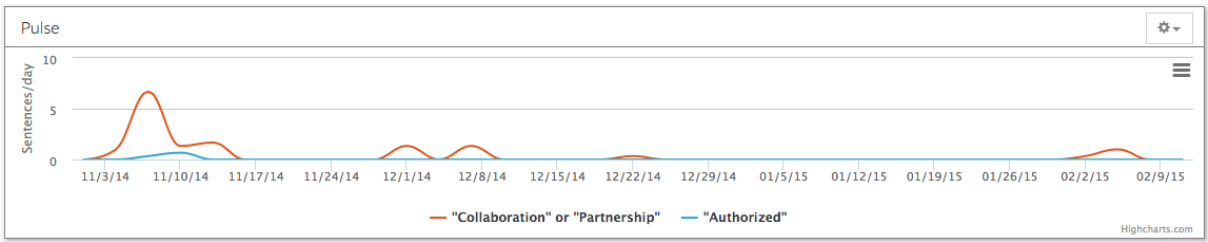

*Sentences about the project that included either a "collaboration" or a "partnership" term compared to sentences that included the word "authorized."*

Such mischaracterizations in the initial coverage of the project likely influenced critiques that focused either on the limited capacity of WAM! to address harassment across Twitter or on "censorship" concerns.[74]

## TIMESERIES ANALYSIS OF COVERAGE

While WAM! was frequently mentioned in the initial coverage spike related to the harassment reporting project, stories in the second spike (2 December, following Twitter's announcement of changes in their reporting tool) more often omitted WAM! and credited the pilot either to Twitter alone or to Twitter and an unnamed nonprofit organization. The graph below shows all coverage related to the pilot project in the US, online-first, and tech media sources in Media Cloud's database. Sentences that refer to the project without mentioning WAM! by name are shown in orange, while sentences that refer to the project and do mention WAM! are shown in blue.

.............................

74.  Within Media Cloud's dataset, the first mention of "censorship" as related to the reporting project is WAM executive director Jaclyn Friedman's characterization of harassment as a form of censorship in a 7 November 2014 interview published in The Atlantic. The first characterization of the project itself as "censorship" or "policing speech" is on 10 November 2014 from blogger Andrew Sullivan, who remained the primary source of censorship-related critique. Articles that mention Sullivan's critiques, however, are overwhelmingly critical of Sullivan himself rather than of the reporting project.

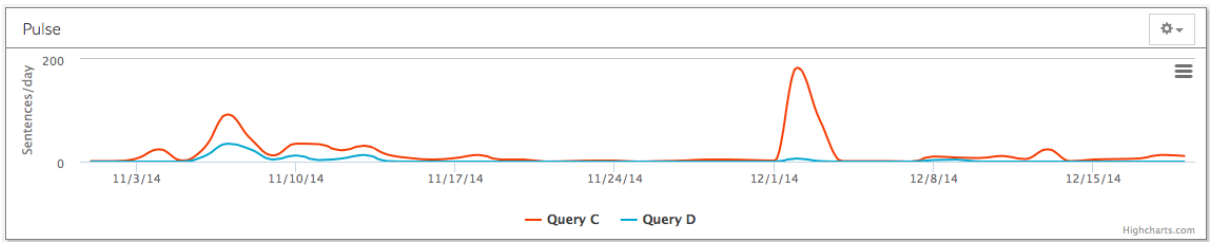

*Sentences per day* mentioning the harassment-reporting project that reference WAM! by name (blue) compared to mentions of the harassment-reporting project overall (shown in orange). [75]

## SETTING THE PROJECT IN CONTEXT: 2014–2015 MEDIA COVERAGE OF ONLINE HARASSMENT

We also used Media Cloud to examine coverage related to harassment or abuse on Twitter more generally in US, online-first, and tech media sources during the thirteen-month period from 1 January 2014 to 15 February 2015. We identified five distinct periods during which attention to the Twitter abuse/harassment issue spiked, two of which were related to the harassment reporting project.

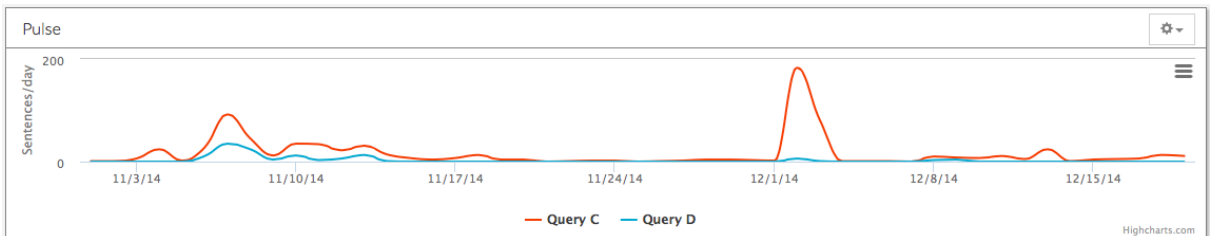

*Sentences per day that mentioned harassment or abuse on Twitter, 1 Jan 2014 to 15 Feb 2015*[7677]

Peaks include:

■ **First Peak: January 2014** peaks are two UK-based events:

▶ 22 Jan 2014: After retired English footballer Stan Collymore comments about online abuse, he receives harassing comments when an ex-girlfriend accuses him of domestic abuse

▶ 24 Jan 2014 is the sentencing of a man & woman who harassed Caroline Criado-Perez.

■ **Second Peak: August 2014**

13-14 Aug 2014: Zelda Williams leaves Twitter after abuse following the death of her father Robin Williams. This sparked a wider conversation about how Twitter handles abuse.

75. This graph is based on analysis of individual sentences about the project, rather than on all of the sentences in each story. Because a story can include both sentences about the project that do mention WAM! by name and sentences about the project that only mention Twitter, some individual stories are represented by both the orange and the blue lines.

76. Mentions of harassment or abuse on Twitter comprise only a small portion of mentions of Twitter overall (no more than 2% of all mentions per day between 1 January 2014 and 15 February 2015. This is due to the frequency with which Twitter is mentioned in stories that are not about Twitter itself—for example, when a spokesperson confirms or denies something via Twitter, or when readers are invited to follow an article's author on Twitter.

77. The January 8 2015 NPR segment about online harassment was not associated with a peak in coverage.

■ **Third Peak: Nov 2014**

▶ 2-5 Nov 2014: Yaya Toure, a soccer player for Manchester City, receives racial abuse after rejoining Twitter

▶ **7 Nov 2014 onward** is associated with **WAM!'s authorized reporting status**

■ **Fourth Peak: 2 December 2014**. Twitter releases its updated harassment reporting system. Coverage is driven by talk of Twitter's efforts to combat harassment, with fewer explicit mentions of WAM of but many mentions of their "collaboration" with a non-profit.

■ **Fifth Peak: 5 Feb 2015**: A Twitter employee posts a Lindy West piece published in the *Guardian* on an internal Twitter forum; Costolo responds with internal memo; internal memo is published by the *Verge*, leading to widespread discussion. This is the only story between 1 Jan 2014 to 15 Feb 2015 that has coverage which exceeds the story on WAM!'s authorized reporting status.

# APPENDIX 3: WAM! REPORTING FORM

The design of the WAM! reporting form, accessible only through the WAM! website, shaped the data the project received and consequently the analysis contained in this document. Though the actual form has since been removed, screenshots of the backup, privately mirrored version of the form follow. These are expanded through written descriptions to detail the possible variations, nonvisible text, and answer restrictions reporters would have encountered.

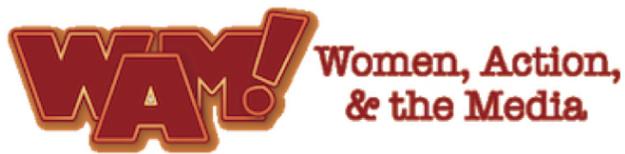

## WAM Twitter Harassment Reporting Tool (Pilot)

This form is to report Twitter harassment to Women Action and the Media for escalation.

We are reviewing submissions and have a goal to respond to requests within 24 hours for more information or to notify you we are escalating to Twitter.

Please fill out the form completely so we can assist you as quickly as possible.

**Are you the person being targeted on Twitter?** *
- ◉ Yes
- ○ No, I am reporting on behalf of someone else

**First and Last Name of the person being harassed**

| | |
|---|---|
| First | Last |

**Email of the person being harassed** *

**Enter Twitter handle being harassed (Do not include the @ symbol)** *

**What date did the harassment start?**

MM / DD / YYYY

**Do you fear for your personal safety due to this harassment?** *
- ○ Yes – I fear my personal safety
- ○ No – I do not fear my personal safety

**How many times have you reported this harassment to Twitter?** *

**Are you being harassed by a single person or by multiple people/accounts?** *

○ Harassed by single person

○ Harassed by multiple people or accounts

○ Unsure

**Please describe in detail the harassment you are receiving.** *

Maximum Allowed: **1000** words.    *Currently Used:* **0** *words.*

**Give an example of the harassing tweets here**

http://twitter.com/

**Please select a category for the type of harassment you are reporting** *

○ Impersonation

○ Threats of violence

○ Doxxing (Releasing your private information)

○ Posting false information (fake quotes attributed to you, altered images, libel)

○ Hate speech (sexist, racist, homophobic, etc.)

○ Encouraging people to harass you via phone or other offline methods

○ Revenge porn or Nonconsensual photography

○ Other

[                              ]

**List the Twitter handles of the people harassing you (or the
targeted person if you are filling this out on their behalf)** *

[                                                              ]
[                                                              ]
[                                                              ]
[                                                              ]

Maximum Allowed: **500** characters.    *Currently Used:* **0** *characters.*

**Are you being harassed on multiple platforms?** *
◉ No
○ Yes

**How many weeks has this harassment been occurring?**

[                              ]

Enter a number between **0** and **500**.

**(Optional) What is your Phone Number?**

[         ] – [         ] – [           ]
  ###         ###          ####

**Is there anything else you would like us to know?**

[                                                              ]

Maximum Allowed: **35** words.    *Currently Used:* **0** *words.*

[ Submit ]

People who visited the WAM! reporting form found a single page of 16–20 questions. The range
reflects the presence of two choice points in the form, where selections could lead to additional
questions. The first set of additional questions opened in conjunction with the very first question: *Are
you the person being targeted on Twitter?* The default response reporters saw was 'Yes.' If reporters
responded 'No,' they were asked to include their own name and email address, as well as indicate
whether the report was being made with the awareness of the target of the harassment. The second
set of additional questions opened in response to the answer to question 13 of the base form: *Are you
being harassed on multiple platforms?* This defaulted to 'No.' If reporters responded 'Yes,' they were
asked additional questions about what other platforms harassment was occurring on.

Multiple questions included rollover explanations, or additional information that became visible
when the reporter's cursor hovered over the question. These served several purposes. Rollovers
were used to provide guidance and clarification. In particular, they were used in conjunction with questions asking for names and emails to explain how to answer these questions if the reporter was not the target of the harassment. Similarly, a rollover was used to explain that a reporter need provide only a single example tweet; more could be shared at a later point. Rollovers were also used to direct the format of input: a rollover associated with the question about the Twitter handle of the target of harassment asked reporters not to include the @ symbol in their answer; a rollover requesting a numerical entry accompanied the question about how many weeks harassment had been occurring. Rollovers were used twice specifically to explain why questions were being asked: for bystanders, to explain that WAM! wanted to analyze data about bystander reporting and the harassment target's awareness of it; and with regard to the phone number request, to explain this might be used for verification or communication. Rollovers were not used with the categories of harassment or the question, *Do you fear for your personal safety due to this harassment?*

Just over half of the questions were required; these are marked with a red superscript star. If any of these were left unanswered, form submission would not complete: reporters would be presented with the form again with missing information brought to their attention. Note that in addition to items marked as required and items left unmarked, the question soliciting the reporter's phone number is specifically labeled "Optional." The format of acceptable answers varied with the question; formats used included radio buttons (the circular answer options) with pre-assigned answers, free text boxes, and numerical entry boxes. Radio buttons allowed only a single selection.